\documentclass[a4paper,11pt]{article}
\pdfoutput=1 

\usepackage{jheppub}
\usepackage{bm,amssymb,slashed,graphicx,multirow,soul,mathtools,xspace,array}  
\usepackage{float}                   
\allowdisplaybreaks
\usepackage{ bbold }
\usepackage{subfigure}

\usepackage{hyperref}
\usepackage[normalem]{ulem}

\usepackage{braket}
\usepackage{booktabs}
\usepackage{multicol}
\usepackage{tabularx}
\usepackage{multirow}

\hypersetup{
  colorlinks=true,
  linkcolor=red,
  citecolor=blue,
  urlcolor=blue,
 hypertexnames=false 
}

\newcommand{\dNeff}{\Delta N_{\text{eff}}}

\newcommand{\eV}{{\, \rm eV}}
\newcommand{\GeV}{{\, \rm GeV}}
\newcommand{\TeV}{{\, \rm TeV}}
\newcommand{\eps}{\epsilon}
\newcommand{\eq}{\text{eq}}

\begin{document}

\preprint{TTP24-002, P3H-24-012}

\title{\bf{Thermal production of astrophobic axions}}

\author[1]{Marcin Badziak}
\author[2,3,4]{Keisuke Harigaya}
\author[1]{Micha\l\ \L ukawski} 
\author[5]{Robert Ziegler} 
\affiliation[1]{Institute of Theoretical Physics, Faculty of Physics, University of Warsaw, ul.~Pasteura 5, PL--02--093 Warsaw, Poland}
\affiliation[2]{Department of Physics, University of Chicago, Chicago, IL 60637, USA}
\affiliation[3]{Enrico Fermi Institute and Kavli Institute for Cosmological Physics, University of Chicago, Chicago, IL 60637, USA}
\affiliation[4]{Kavli Institute for the Physics and Mathematics of the Universe (WPI),
        The University of Tokyo Institutes for Advanced Study,
        The University of Tokyo, Kashiwa, Chiba 277-8583, Japan}
\affiliation[5]{
	Institut f\"ur Theoretische Teilchenphysik, Karlsruhe Institute of Technology, Karlsruhe, Germany}
\date{}

\abstract{
Hot axions are produced in the early Universe via their interactions with Standard Model particles, contributing to dark radiation commonly parameterized as $\dNeff$. In standard QCD axion benchmark models, this contribution to $\dNeff$ is negligible after taking into account astrophysical limits such as the SN1987A bound.
We therefore compute the axion contribution to $\dNeff$ in so-called astrophobic axion models characterized by strongly suppressed axion couplings to nucleons and electrons, in which astrophysical constraints are relaxed and $\dNeff$ may be sizable. We also construct new astrophobic models in which axion couplings to photons and/or muons are  suppressed as well, allowing for axion masses as large as few eV.
Most astrophobic models are within the reach of CMB-S4, while some  allow for $\dNeff$ as large as the current upper bound from Planck and thus will be probed by the Simons Observatory. The majority of astrophobic axion models predicting large $\dNeff$ is also within the reach of IAXO or  even BabyIAXO. 

}

\maketitle

\section{Introduction}

Among the most compelling solutions to the strong CP problem is the Peccei-Quinn (PQ) mechanism~\cite{Peccei:1977hh,Peccei:1977ur} that predicts the existence of a new light pseudoscalar particle called the QCD axion~\cite{Weinberg:1977ma,Wilczek:1977pj}, which is also an excellent candidate for cold dark matter (DM). Most of the parameter space in which the axion can explain the observed DM abundance has not been probed experimentally, but several experiments targeting the most interesting axion region are in preparation, see Ref.~\cite{DiLuzio:2020wdo} for a review.

To successfully account for cold dark matter, QCD axions have to be produced non-thermally in the early universe~\cite{AxionDM1, AxionDM2, AxionDM3}. However, QCD axions could also be thermally produced via their interactions with the Standard Model (SM) bath, and contribute to the energy density of relativistic degrees of freedom, usually  parametrized as the effective number of neutrinos $N_{\rm eff}$. This axion contribution, dubbed in the following $\dNeff$,  is constrained by observations of the cosmic microwave background (CMB) and low-redshift baryon acoustic oscillations (BAO) data. The most recent (2018) analysis from the Planck collaboration provides the combined constraint $\dNeff \le 0.3$ at the 95\% confidence level (CL)~\cite{Planck:2018vyg}. 
In the near future this bound may be further improved by a factor of few with the help of the Simons Observatory, down to about 0.1~\cite{SimonsObservatory:2018koc}, and eventually by the CMB-S4 experiments down to about 0.05~\cite{CMB-S4:2016ple}.  

The axion couples to the SM particles with strength inversely proportional to the axion decay constant $f_a$. For sufficiently small $f_a$, the axion is in thermal equilibrium with the SM bath for some range of temperatures. As the temperature drops, axion interactions become suppressed, and eventually freeze out at a temperature $T_d$. Smaller $f_a$ leads to later axion decoupling from the SM bath and hence larger $\dNeff$, since $\dNeff \propto g_{*s}(T_d)^{-4/3}$ (assuming instantaneous decoupling), where $g_{*s} (T_d)$ is the effective number of SM entropy degrees of freedom at $T_d$.  On the other hand, for sufficiently large $f_a$, the axion is never in thermal equilibrium and only produced via its interactions with the SM bath through thermal freeze-in~\cite{Hall:2009bx}. In this case the late-time abundance is proportional to the production rates scaling as $\propto 1/f_a^2$, so that $\dNeff \propto f_a^{-8/3}$. 

Using the current constraint\footnote{For sufficiently small $f_a$ the axion is no longer relativistic at the time of recombination, and its contribution to the total energy density is constrained rather by limits on warm DM, although one can still formulate this bound in terms of $\dNeff$~\cite{Caloni:2022uya}.} on $\dNeff$, it is therefore possible to set a lower bound on $f_a$, which can be further improved in the near future. The lower bound on $f_a$ obtained from $\dNeff$ depends on the nature of the axion interactions, and thus is model-dependent. For example, in the KSVZ model~\cite{Kim:1979if,Shifman:1979if} the current bound is $f_a\gtrsim2\times10^7\GeV$ and is expected to be improved up to $f_a\gtrsim6\times10^7\GeV$ with CMB-S4 data~\cite{Notari:2022ffe} (see also Refs.~\cite{DEramo:2021lgb,Bianchini:2023ubu}). However, smaller $f_a$ also leads to more efficient axion emission in various stellar environments, and the resulting bounds are typically much stronger than those from $\dNeff$. In the KSVZ model, the cooling rate of neutron stars (NS)~\cite{Buschmann:2021juv} and the duration of the neutrino burst from SN1987A~\cite{Carenza:2019pxu} lead, independently, to a lower bound on $f_a$ of about $4\times10^8\GeV$. For such large $f_a$, the KSVZ model predicts $\dNeff\lesssim0.03$ so it is difficult to probe the KSVZ axion with cosmological data in the foreseeable future. The same conclusion is valid for simple DFSZ models~\cite{Zhitnitsky:1980tq,Dine:1981rt}, which have been studied in the context of thermal axion production in e.g. Refs.~\cite{Ferreira:2018vjj,Arias-Aragon:2020shv,Ferreira:2020bpb,DEramo:2021lgb,DEramo:2022nvb}.

For this reason future experimental probes of $\dNeff$ are only sensitive to QCD axion models in which either astrophysical bounds are absent for some reason or the relevant axion couplings are suppressed. For example, in order to relax the constraints on $f_a$ from  NS cooling and SN1987A, it is necessary to suppress simultaneously the axion couplings to protons and neutrons. This may be possible if the axion couples to up- and down-quarks in such a way that the model-independent contribution to axion-nucleon couplings from the QCD anomaly is cancelled to good approximation. Also axion couplings to electrons need to be suppressed, in order to avoid stringent constraints from white dwarfs~\cite{MillerBertolami:2014rka}. Such models, dubbed  ``astrophobic" axion models, have been discussed in Refs.~\cite{DiLuzio:2017ogq, Bjorkeroth:2018ipq, Bjorkeroth:2019jtx, Badziak:2021apn, DiLuzio:2022tyc, Badziak:2023fsc,Takahashi:2023vhv}.  The goal of the present paper is to systematically compute $\dNeff$ in these kind of models, which  allow to satisfy astrophysical constraints with rather low $f_a$, and assess their prospects for discovery through $\dNeff$ by the Simons Observatory~\cite{SimonsObservatory:2018koc} and CMB-S4~\cite{CMB-S4:2016ple}, laboratory searches such as IAXO~\cite{IAXO:2020wwp}, and the James Webb Space Telescope (JWST)~\cite{Roy:2023omw}. 

In common QCD axion benchmark models the dominant contribution to $\dNeff$ comes from axion scatterings with pions ($\pi\pi\leftrightarrow\pi a$)~\cite{Chang:1993gm}. However, this production mode is necessarily suppressed in astrophobic axion models, because the suppression of axion-nucleon couplings also leads to the suppression of axion-pion couplings. We will show, in a model-independent way, that other contributions to $\dNeff$  can still be sizeable, even after taking into account all astrophysical constraints. We will first discuss  the original astrophobic models~\cite{DiLuzio:2017ogq}, which are two-Higgs-doublet models (2HDM) that generalize common DFSZ scenarios. By allowing for flavor non-universal PQ charges, axion couplings to nucleons and electrons can be suppressed, so that in these scenarios axions are mainly produced from lepton flavor-violating (LFV) decays $\tau\to e a$, which are unavoidable in this class of models in order to suppress the axion-electron coupling. This leads to a sharp prediction for $\dNeff$ that only depends on $f_a$, which is however limited by astrophysical constraints on the axion-photon coupling and axion-muon coupling.

For this reason we consider ``proper" astrophobic models, in which not only axion couplings to nucleons and electrons, but also to muons and/or photons are suppressed. Examples of such models have been recently proposed in Ref.~\cite{Badziak:2023fsc}, in which the coupling of the axion with SM particles are controlled by the PQ charges of SM fermions and astrophobia is naturally obtained without any tuning. See Ref.~\cite{Lucente:2022vuo} for earlier attempts.
We also construct new ``proper" astrophobic models and show that the required suppression of couplings can be realized in generalized DFSZ models with three Higgs doublets (3HDMs), and systematically classify such models, which not only allow for axion decay constants $f_a<10^7\GeV$ compatible with all constraints, but also to suppress the axion-electron coupling without tuning (in contrast to 2HDMs).

The rest of the article is organized as follows. In Section~\ref{couplings} we discuss the general QCD axion effective Lagrangian and summarise laboratory and astrophysical constraints on the axion decay constant. In Section~\ref{sec:MI} we analyse dominant channels for thermal axion production in a model independent way. In Section~\ref{sec:2HDM} we compute the thermal axion abundance in astrophobic 2HDMs, while in Section~\ref{sec:3HDM} we present a similar analysis for 3HDMs. 
In Section~\ref{sec:natural} we discuss cosmological constraints on the naturally astrophobic axion. 
We conclude our work in Section~\ref{sec:concl}, which is followed by several appendices, in which we present the details of the computations of axion couplings to pions and kaons in Chiral Perturbation Theory (Appendix~\ref{chiPT}), thermal axion production rates (Appendix~\ref{rates}), approximate solutions to the Boltzmann equation (Appendix~\ref{Boltzmann}) and the explicit construction of astrophobic models (Appendix~\ref{models}).

\section{QCD Axion  Couplings}  
\label{couplings}
In this section we define the general effective Lagrangian and review the laboratory and astrophysical constraints on axion couplings relevant for our analysis. 
\subsection{Axion Effective Lagrangian}

At energies much below the PQ breaking scale, the effective axion couplings to gauge fields and fermions are given by
\begin{equation}
{\cal L}  =  \frac{a}{f_a} \frac{\alpha_s}{8 \pi}  G \tilde{G} +  \frac{E}{N}  \frac{a}{f_a} \frac{\alpha_{\rm em}}{8 \pi} F \tilde{F} + \frac{\partial_\mu a}{2 f_a} \overline{f}_i \gamma^\mu \left( C^V_{ij} + C^A_{ij} \gamma_5 \right) f_j \, ,
\label{LaPQ}
\end{equation}
where $f_a$ is the axion decay constant, $F \tilde{F} \equiv 1/2 \, \eps_{\mu \nu \rho \sigma} F^{\mu \nu} F^{\rho \sigma}$ with the electromagnetic (EM) field strengths $F^{\mu\nu}$ and similar for gluons, $E/N$ is the ratio of EM and color anomaly coefficients and we use the convention $\eps^{0123} = -1$. For later convenience we define $C_{ij} = \sqrt{|C_{ij}^A|^2 + |C_{ij}^V|^2}$ for the flavor-violating couplings, as thermal axion production typically does not depend on the chiral structure of axion couplings. 

The first term in Eq.~\eqref{LaPQ} gives rise to the axion mass, which can be conveniently calculated in chiral perturbation theory, giving~\cite{Gorghetto:2018ocs} 
\begin{align}
m_a & = 0.5691(51) \,  {\rm eV} \left( \frac{10^{7} \GeV}{f_a} \right)  \, .
\end{align}
Below the scale of the QCD phase transition the relevant couplings are those to photons, nucleons, leptons and pions, 
\begin{align}
{\cal L} & =    C_\gamma \frac{a}{f_a} \frac{\alpha_{\rm em}}{8 \pi} F \tilde{F}  
 +\frac{\partial_\mu a}{2 f_a}  C_N \overline{N} \gamma^\mu \gamma_5  N  +\frac{\partial_\mu a}{2 f_a}   \overline{\ell}_i \gamma^\mu \left( C^V_{ij} + C^A_{ij} \gamma_5 \right) \ell_j 
 + \frac{ \partial_\mu a}{f_a f_\pi} C_\pi \partial [\pi\pi\pi]_\mu
 \, ,
\label{LaQCD}
\end{align}
where $N = n,p$ and $\partial [\pi\pi\pi]_\mu=2\partial_\mu \pi_0 \pi_+ \pi_-  -   \pi_0 \partial_\mu \pi_+ \pi_-  -   \pi_0  \pi_+ \partial_\mu \pi_-$. Matching to the UV coefficients in the Lagrangian of Eq.~\eqref{LaPQ} gives~\cite{Badziak:2023fsc}
\begin{align} 
\label{CppCn}
C_p +C_n &= 
0.40(4)\left(0.95\left(C_u + C_d\right) + 0.05 -\frac{1+z}{1+z+w}\right) - 2 \delta  \, ,  \\
\label{CpmCn}
C_p -C_n &= 1.273(2)\left( C_u  - C_d  - {\frac{1-z}{1+z+w}} \right) \, , \\
\label{Cpi}
C_\pi & = - \frac13 \left( C_u  - C_d  - {\frac{1-z}{1+z+w}} \right) \, , \\
\label{Cgam}
C_{\gamma} &= 2 \pi f_a g_{a \gamma \gamma} = E/N - 2.07(4)   \, ,  
\end{align}
where $C_q \equiv C^A_{qq} (\mu = f_a)$, $z = m_u/m_d = 0.48(2)$, $w=m_u/m_s=0.023(1)$, and $\delta \approx \sum_{i=s,c,b,t}{\delta_i}C_i$. The coefficients $\delta_i$ for $i=s,c,b$  are $\mathcal{O}(10^{-2})$, arising from QCD renormalization group (RG) effects, and their exact values can be found in Ref.~\cite{Badziak:2023fsc}. On the other hand, $\delta_t\sim\mathcal{O}(0.1)$  is much larger due to RG effects induced by the large top Yukawa coupling~\cite{Choi:2017gpf,Chala:2020wvs,Bauer:2020jbp,Choi:2021kuy}, and its exact value is sensitive to $f_a$ and the details of the UV model.
 We also note that $C_\pi\propto \left(C_n-C_p \right)$, so that pion couplings are suppressed whenever couplings to nucleons are suppressed. In the above result for $C_{\gamma}$ we quote the value $2.07(4)$ obtained in Ref.~\cite{Badziak:2023fsc} using the results from Ref.~\cite{Lu:2020rhp}, which include the effects of the strange quark within three-flavor ChPT at NLO. This value is different from the usually quoted value $1.92(4)$ from Ref.~\cite{GrillidiCortona:2015jxo}. Nevertheless, this difference does not have important impact on our results. 

\subsection{Constraints on Axion Couplings}\label{subsection:couplingsConstraints}

There are several constraints on axion couplings from astrophysics and laboratory searches. In the following we collect the limits on axion couplings to nucleons, pions, electrons, muons, photons and LFV couplings, which are most relevant for our analysis.

\subsubsection{Nucleon Couplings}

Since the axion-gluon coupling is crucial in solving the strong CP problem with the PQ mechanism, axion-nucleon couplings are generically present and mainly constrained by astrophysical observations, prominently the duration of the neutrino burst observed in SN1987A and the cooling rate of neutron stars. The constraints obtained from SN1987A~\cite{Carenza:2019pxu} and neutron stars~\cite{Buschmann:2021juv} are roughly comparable, but for concreteness we only use the SN1987A bound from Ref.~\cite{Carenza:2019pxu}, which provides a formula for the general case when the axion couples differently to neutrons and protons:
\begin{equation}
0.61 g_{ap}^2+g_{an}^2 + 0.53 g_{an} g_{ap} < 8.26\times 10^{-19} \, , 
\end{equation}
where $g_{ai}\equiv C_{i} m_{i}/f_a$ with $i=n,p$. For $C_n = 0$ this leads to the lower bound 
\begin{equation}
\label{bound_nucleon}
\frac{f_a}{|C_p|} \gtrsim 8 \times 10^8\GeV \, , 
\end{equation}
while varying $C_n$ in the range $|C_n| \le |C_p|$ can strengthen the above bound by at most a factor of 2. In the KSVZ model, $C_p\approx-0.47$, which implies $f_a \gtrsim 4\times10^8\GeV$. As we will see in the next section, large values of $\dNeff$ typically require much smaller values of $f_a$, and thus suppressed nucleon couplings $|C_p|,|C_n|\lesssim\mathcal{O}(10^{-2})$. Indeed axion-nucleon couplings can be suppressed if there is an approximate cancellation between the axion-gluon and axion-quark contributions~\cite{DiLuzio:2017ogq}. This happens if $C_u\approx2/3$ and  $C_d\approx1/3$, as can be seen from Eqs.~\eqref{CppCn} and \eqref{CpmCn}. We refer to axions satisfying these criteria as ``nucleophobic" axions. Due to higher order corrections to the axion-nucleon couplings, the values of $f_a$ satisfying the bounds from SN1987A and neutron stars cannot be arbitrary low, but values as small as $10^7$~GeV (or even $10^6$~GeV if $z$ is very close to 0.49) may be allowed~\cite{Badziak:2023fsc}.

\subsubsection{Pion Couplings} 

Suppressed couplings to nucleons imply suppressed pion couplings. The maximal pion coupling consistent with the constraints on the nucleon couplings is obtained for $C_p \approx -C_n$. Using Eqs.~\eqref{CpmCn} and \eqref{Cpi} one can relate  pion couplings to nucleon couplings
\begin{equation}
C_\pi\approx- \frac14(C_p-C_n) \approx \frac12C_n \,.
\end{equation} 
Using the bound on the neutron coupling \eqref{bound_nucleon} one can derive an upper bound on the pion coupling to the astrophobic axion:
\begin{equation}
C_\pi\lesssim0.5\frac{f_a}{10^9\GeV} \,.
\end{equation} 
As we will see below, for an axion-pion coupling satisfying the above constraint, $\dNeff$ from axion-pion scattering is below 0.01, and thus negligible given near future sensitivities.

\subsubsection{Electron Coupling} 

The axion-electron coupling is constrained by the observed shape of the white dwarf luminosity function, giving the 95\% CL lower  bound~\cite{MillerBertolami:2014rka} 
\begin{equation}
\frac{f_a}{|C_e|} \gtrsim 2\times 10^9\GeV \, .
\end{equation}
This implies that the axion-electron coupling must also be suppressed, at least to the level $\mathcal{O}(10^{-2})$ in order to allow for sizeable contributions to $\dNeff$.

\subsubsection{Muon Coupling} 

Also the axion-muon coupling is constrained by the energy-loss argument for SN1987A~\cite{Bollig:2020xdr,Croon:2020lrf, Caputo:2021rux}, which lead to the conservative lower bound~\cite{Caputo:2021rux}

\begin{equation}
\frac{f_a}{|C_\mu|} \gtrsim  1.2\times 10^{7}\GeV \, .
\end{equation}
For ${\cal O}(1)$ couplings the limit on $f_a$ from the axion-muon coupling is much weaker than the bounds on nucleon and electron couplings. However, in nucleophobic and electrophobic axion models, this becomes a relevant constraint and limits the maximal contribution to $\dNeff$ from axion scatterings with muons.

\subsubsection{Photon Coupling} 

Observations of the evolution of horizontal branch stars in globular clusters constrain the axion-photon coupling as~\cite{Ayala:2014pea} 
\begin{align}
\frac{f_a}{ \left| C_\gamma \right|} > 1.8 \times 10^7 ~{\rm GeV}  \,.
\end{align}
In order to satisfy the above bound, $f_a$ should be above $\mathcal{O}(10^7) \GeV$,  unless the axion-photon coupling is suppressed. This is indeed the case for $E/N=2$, as noted already in Ref.~\cite{Kaplan:1985dv}, and $f_a$ down to $\mathcal{O}(10^6) \GeV$ is allowed. Note also that within theoretical uncertainties even $C_\gamma=0$ is possible for $E/N=2$.

\subsubsection{Flavor-violating Couplings} 

Flavor-violating axion couplings are constrained by high-intensity laboratory experiments looking for missing energy in rare decays.
The strongest constraint is set by the experiments searching for $\mu \to e a$ decays at TRIUMF~\cite{Jodidio:1986mz} or TWIST~\cite{TWIST:2014ymv}, depending on the chirality structure of the axion couplings. For purely left-handed or right-handed couplings the upper bounds at $95\%$ CL read~\cite{Calibbi:2020jvd}
\begin{align}
\frac{f_a}{|C_{\mu e}|} & \ge 5.0 \times 10^8 \GeV\, ,   & {\rm for}\ C_{\mu e}^V & = - C_{\mu e}^A  \,, & \\
\frac{f_a}{|C_{\mu e}|} & \ge 2.5 \times 10^9 \GeV\, ,   & {\rm for}\ C_{\mu e}^V & = C_{\mu e}^A  \,.
\end{align}
For values of $f_a$ where $\dNeff$ can be non-negligible, the above limits imply $|C_{\mu e}|\ll1$.

Constraints from flavor-violating $\tau$-decays are much weaker. The strongest constraints have been recently provided by the Belle-II~\cite{Belle-II:2022heu} collabaration, which result in the following lower bounds at $95\%$ CL 
\begin{align}
\frac{f_a}{|C_{\tau e}|} & \gtrsim  3.6\times 10^6\GeV  \, , \\
	\frac{f_a}{|C_{\tau \mu|}} & \gtrsim  4.6\times 10^6\GeV  \,.
\end{align}
Rescaling the current expected bounds provided in Ref.~\cite{Belle-II:2022heu} for 62.8 fb$^{-1}$, Belle-II with 50 ab$^{-1}$ can be expected to probe flavor-violating $\tau$-couplings down to
 \begin{align}
 \label{Belle2}
	\frac{f_a}{|C_{\tau e}|} & \gtrsim  1.6\times 10^7\GeV \,,  \\
	\frac{f_a}{|C_{\tau \mu|}} & \gtrsim 1.7\times 10^7\GeV \,.
\end{align}

\section{Model-independent Analysis of $\dNeff$}
\label{sec:MI}
We first perform the calculation of $\dNeff$ in nucleophobic axion models in a model independent way, allowing all lepton-axion couplings that are consistent with experimental and astrophysical constraints\footnote{Model-independent  analyses of thermal axion production in various channels without taking into account astrophysical constraints were presented  in Refs.~\cite{DEramo:2018vss,Green:2021hjh,DEramo:2021usm}.}. This will allow us to understand which couplings are the most relevant for thermal axion production in astrophobic models. We do not study the impact of axion-quark couplings on $\dNeff$ in this section, as they cannot be reliably computed  due to non-perturbative effects, but will discuss their potential impact in the following sections for specific models.  

\begin{figure}[t]
\begin{center}
 \includegraphics[width=0.88\textwidth]{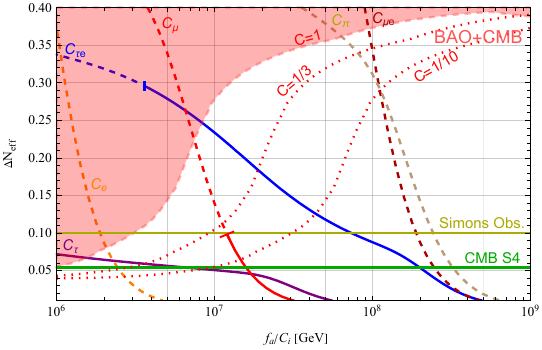}
 \caption{Additional effective number of neutrinos $\dNeff$ as a function of $f_a/C_i$, for $i = \pi, \tau, \mu, e, {\mu e}, {\tau e}$. $C_{\tau \mu}$ gives predictions essentially identical to $C_{\tau e}$. The red region is excluded by BAO+CMB \cite{Caloni:2022uya} assuming $C_i=1$, while dotted curves show the bound for the representative values $C_i=1/3$ and $C_i =1/10$.  Dashed (solid) lines indicate ranges of excluded (viable) values of $f_a/C_i$ for a given type of coupling, as discussed in Section~\ref{subsection:couplingsConstraints}.  
\label{fig:dNeff_MI} }
\end{center}
\end{figure}

A crucial feature of nucleophobic models is that the axion-pion coupling is tiny, since it is proportional to $C_p-C_n$, which in turn must be strongly suppressed to avoid axion couplings to nucleons. In contrast to common axion benchmark models, where $\pi\pi \to \pi a$ scattering is the dominant source of axion thermalization, nucleophobia implies that other processes for axion production become relevant. In Fig.~\ref{fig:dNeff_MI} we show the predicted value of $\dNeff$ as a function of $f_a/C_i$, assuming the presence of a single axion coupling $C_i$ at a time. Details of the calculation are outlined in Appendices \ref{rates} and \ref{Boltzmann}.%
\footnote{For freeze-in production, where the axion does not reach thermal equilibrium, our computation underestimates $\dNeff$. The underestimation is most significant for muon scattering, where it is off by at most a factor 2 in the phenomenologically relevant parameter range. The sensitivity of future CMB observations on $f_a$ is largely unaffected by this, due to the strong dependence of $\dNeff$ on $f_a$ for the freeze-in case. See Appendix \ref{Boltzmann} for details.
}
We restrict to leptonic couplings $C_\tau, C_\mu, C_e,  C_{\mu e}, C_{\tau e}$ ($C_{\tau \mu}$ gives predictions essentially identical to $C_{\tau e}$), and also include $C_\pi$  for comparison. 
The constraints on $f_a/C_i$ discussed in Section~\ref{couplings} are taken into account by dashing the predicted curve $\dNeff$ for values of $f_a/C_i$ below the respective limit, while drawing it solid where the bound is respected.

Present CMB and BAO data exclude certain regions in the $(m_a, \dNeff)$ plane~\cite{Caloni:2022uya}, which one can convert to limits in the $(f_a/C_i, \dNeff)$ plane using the QCD axion mass relation and fixing $C_i$. We show in red the region excluded for $C_i = 1$, and show the contours of the excluded regions for $C_i = 1/3$ and $C_i = 1/10$.  While the bounds on the axion couplings are essentially independent of $C_i$ for large $f_a$, they drastically strengthen for $f_a<\mathcal{O}(10^7) \GeV$, corresponding to $m_a>\mathcal{O}(0.6) \eV$, where the axion can no longer be treated as massless and becomes non-relativistic around recombination.
Such axions affect the CMB in a different way than extra relativistic degrees of freedom. In Fig.~\ref{fig:dNeff_MI} we take this into account by using the bounds on $\dNeff$ presented in Ref.~\cite{Caloni:2022uya}, which have been obtained for general axion masses.%
\footnote{Note that for sufficiently heavy axions $\dNeff$ no longer denotes extra relativistic degrees of freedom, but remains a useful parametrization of the total energy density of thermally produced axions.}
Similar effects will be relevant for the expected limits from the Simons Observatory and CMB-S4. However, in order to show the sensitivity of future experiments in Fig.~\ref{fig:dNeff_MI}, we conservatively assume forecasted sensitivities to $\dNeff$ for massless axions $\dNeff^{\text{Simons}}(2\sigma) = 0.1$ \cite{SimonsObservatory:2018koc} and $\dNeff^{\text{CMB-S4}} (2\sigma) = 0.054$ \cite{CMB-S4:2016ple}.

It is clear from Fig.~\ref{fig:dNeff_MI}  that for generic (flavor-universal) models, where all axion couplings are of similar size, $\pi\pi \to \pi a$ scattering gives by far the largest contribution to $\dNeff$. In hadronic axion models, such as KSVZ with $C_\pi\approx0.1$, this results in the so-called hot dark matter bound $f_a\gtrsim2\times 10^7 \GeV$, which is expected to be improved up to $f_a\gtrsim6\times 10^7 \GeV$ by CMB-S4 experiments~\cite{Notari:2022ffe}. However, the astrophysical constraints require $f_a/C_\pi \gtrsim 2 \times 10^9 \GeV$ (the entire $C_\pi$ curve is dashed), which implies that for values of $f_a$ consistent with these constraints $\dNeff$ is much below $0.01$. Therefore axion interactions with leptons often lead to larger $\dNeff$ than production from pion scattering, after taking into account the astrophysical constraints on $f_a$. 

We first consider the case in which flavor-violating axion couplings are absent. Axion scatterings with electrons give non-negligible contribution to $\dNeff$ only for $f_a/C_e$ below ${\rm few} \times 10^6 \GeV$ (because the scattering cross-section is suppressed by the small electron mass, cf. Appendix~\ref{rates}), but such small values of $f_a/C_e$ are already excluded by astrophysical constraints.
The contribution to $\dNeff$ from axion-muon scatterings is less limited by astrophysics, but after taking into account the bound from SN1987A, $\dNeff$ from axion production off muons is at most about $0.1$, which is at the verge of the future sensitivity of the Simons Observatory. 
Axion-tau couplings instead are not constrained by astrophysics, but axion production off taus leads to $\dNeff>0.05$ only for $f_a/C_\tau\lesssim 10^7$~GeV. Note that for large $f_a$ the curves indeed follow the expected scaling $\dNeff \propto (m_\ell/f_a^2)^{-4/3}$, cf.~Appendix~\ref{FIapp}, in particular roughly the same $\dNeff \approx 0.01$ is obtained for constant values of $m_\ell/f_a^2$ for all three leptons.

Turning to lepton flavor-violating  axion couplings, it is instead possible to produce sizable values of $\dNeff$ even for large $f_a/C_i$, as can be seen from Fig.~\ref{fig:dNeff_MI}. If all LFV couplings are of similar size, the largest contribution comes from $\mu\to e a$ axion production controlled by $C_{\mu e}$,  but the strong laboratory constraints on these decays limit the resulting contribution to $\dNeff$ to negligible values. Instead LFV decays $\tau \to \ell a$ with $\ell = \mu,e$ that are controlled by $C_{\tau \ell}$ are much less constrained by experiments and can give sizable contributions to $\dNeff$. 
 If such couplings are present, CMB-S4 will be sensitive to values of $f_a/C_{\tau \ell}$ as large as $2\times 10^8$~GeV, which is an order magnitude stronger than the forecasted sensitivity of Belle-II, cf. Eq.~\eqref{Belle2}. The current lower bound on $f_a$ from $\dNeff$ is about $8\times 10^6$~GeV for $C_{\tau \ell}=1$, but it is very sensitive to the actual value of $C_{\tau \ell}$, as in this regime the axion ceases to be a relativistic particle at recombination. Accordingly the bound on $\dNeff$ gets stronger for  $f_a<\mathcal{O}(10^7) \GeV$, and  values of $f_a$ much below $10^7$~GeV are excluded, even for $C_{\tau l}$ significantly below unity. Also axion production from decays roughly follow the expected scaling for large $f_a$, $\dNeff \propto (m_\ell/f_a^2)^{-4/3}$, cf.~Appendix~\ref{FIapp}.

To summarize, Fig.~\ref{fig:dNeff_MI} demonstrates that sizable contributions to $\dNeff$ from axion production in the reach of near-future experiments are possible in models where {\it i}) flavor-violating axion-tau couplings are sizable and 
 {\it  ii}) axion couplings to nucleons and electrons, as well as all other flavor-violating axion couplings, are sufficiently suppressed in order to allow for $f_a \lesssim 10^8 \GeV$. The above requirements can be fulfilled in the ``astrophobic" models proposed in Ref.~\cite{DiLuzio:2017ogq}, which we analyze  in the next sections.

\section{ $\dNeff$ in Astrophobic 2HDMs}
\label{sec:2HDM}

We now discuss explicit models that realize astrophobic axions, i.e., axions with suppressed nucleon and electron couplings. In the previous section we have shown that a significant contribution to $\dNeff$ can arise from sizable LFV axion couplings that are not in conflict with astrophysical and rare-decay constraints.
Interestingly, astrophobic axions obtained in DFSZ-like models with two Higgs doublets necessarily imply PQ charges that are flavor non-universal~\cite{DiLuzio:2017ogq}. In the following we discuss the structure of these models and calculate the resulting axion contribution to $\dNeff$. 

There are four different models with two Higgs doublets (2HDM) that feature potentially nucleophobic QCD axions~\cite{DiLuzio:2017ogq, Badziak:2021apn}. The definition and the details of these scenarios, dubbed Q1-Q4, are given in Appendix \ref{models} and summarized in Table~\ref{Qmodels2}. These models depend on the choice of a single vacuum angle $t_\beta \equiv \tan_\beta$ (bounded by perturbativity), and the unitary rotations describing the transition between interaction and mass basis. They can be conveniently parametrized introducing
\begin{align}
\xi^{f_{P}}_{ij} \equiv (V_{fP})^*_{3i}  (V_{fP})_{3j} \, , 
\end{align}
with $f = u,d,e$, $P= L,R$, which depend on the unitary rotations $(V_{fP})_{ij}$ that diagonalize  quark and charged lepton mass matrices. These parameters  satisfy
 \begin{align}
 0 & \le \xi^{f_P}_{ii} \le 1 \, , & \sum_i \xi^{f_P}_{ii}  & = 1 \, , &  |\xi^{f_P}_{ij}| & = \sqrt{\xi^{f_P}_{ii} \xi^{f_P}_{jj}} \, ,
 \label{xirelations}
 \end{align}
 so that there are only two independent parameters in each fermion sector $f_P$ (ignoring complex phases). This also implies that there is no flavor violation if $\xi^{f_{P}}_{ii} =1$ for some $i = 1,2,3$.

After imposing the condition for nucleophobia, $C_u=2/3$ and $C_d=1/3$, the resulting predictions for the remaining axion-quark couplings differ between the scenarios and we show representative models in Tables~\ref{2HDM:quarkCharges} that have been selected as follows. While for models Q1 and Q4 nucleophobia fixes all couplings, there is still some freedom in models Q2 and Q3, corresponding to the choice of quark flavor rotations.  To avoid stringent bounds from flavor-violating meson decays, we only consider the case that flavor-violation in the quark sector is absent\footnote{This is a conservative assumption as far as $\dNeff$ is concerned, since the presence of flavor-violating couplings can only increase $\dNeff$.}
, which leaves four models in each class (Q2 or Q3), only differing in predictions for 2nd and 3rd generation quark couplings. As these choices do not have much impact on our analysis, we simply choose two representative models for each class that we denote as Q2 and Q3 in Table~\ref{2HDM:quarkCharges} (this choice corresponds to $\xi^{u_L}_{33} = \xi^{d_L}_{33} = 1$ for model Q2, and  $\xi^{u_R}_{22} = \xi^{d_R}_{22} = 1$ for model Q3). 

 \begin{table}[t]
	\centering
	\begin{tabular}{|c||c||c|c||c|c|}
		\hline
		Model & $E_Q/N$   & $C_c$ & $C_s$& $C_t$ & $C_b$    \\ \hline
		Q1 & 14/3  &2/3 & -2/3 & 2/3 &-2/3  \\ \hline 
		Q2 &  8/3 & 2/3 & 1/3 & -1/3 & -2/3 \\ \hline 
		Q3 &  8/3 &  -1/3 & -2/3 & 2/3 & 1/3 \\ \hline 
		Q4 & 2/3 & -1/3 & 1/3 & -1/3 & 1/3  \\ \hline \hline
		Q5 (3HDM) & 2 & 0 & 0 & 0 & 0  \\ \hline
	\end{tabular}
\caption{Quark sectors in astrophobic 2HDM and 3HDM axion models after imposing conditions for nucleophobia,  $C_u=2/3$ and $C_d=1/3$.  For Q2 and Q3 additional choices has been made for quark flavor rotations, as explained in the text. Q5 can be realized only in 3HDMs. \label{2HDM:quarkCharges}}
\end{table}

\begin{table}[t]
	\centering
	\begin{tabular}{|c||c||c|c||c|}
		\hline
		Model & $E_L/N$   & $C_\mu$ & $C_\tau$& $C_{\tau e}$    \\ \hline
		E1 & -2  &-2/3 & -1/3 &   $2/3$ \\ \cline{1-1} \cline{3-5}
		\hline 
		E2 &  0 &1/3 & -1/3 &  $2/3$ \\ \cline{1-1} \cline{3-5}
		\hline 
	\end{tabular}
\caption{Lepton sectors in astrophobic 2HDM axion models after imposing conditions for nucleophobia,  $C_u=2/3$ and $C_d=1/3$, electrophobia $C_e=0$ and requiring $C_{\mu e}=0$ in order to avoid stringent constraints from $\mu\to e a$ decays. 
}
\label{2HDM:leptonCharges}
\end{table}

In order to satisfy the astrophysical bounds on the axion-electron coupling for $f_a\lesssim10^8$~GeV as needed for sizable $\dNeff$, the axion must also be electrophobic to good approximation. As discussed in Appendix \ref{models}, within 2HDM there are essentially two potentially electrophobic scenarios\footnote{There are four different models, but they differ only pairwise by the chiral structure of LFV couplings, which is irrelevant for our analysis.}, 
dubbed E1 and E2, which are summarized in Table~\ref{Lmodels2}. Electrophobia ($C_e=0$) is achieved through a tuning of flavor rotations, $\xi_{11}^{e_L} \approx c_\beta^2 (s_\beta^2 )$ for E1 (E2), and since nucleophobia requires $c_\beta^2 = 2/3$, there is necessarily lepton-flavor violation since $\xi_{11}^{e_L} \ne 0, \xi_{11}^{e_L}  \ne1$. In order to avoid strong constraints from $\mu\to e a$ decays, flavor violation cannot be present in the $\mu-e$ sector, which implies $\xi_{22}^{e_L} \approx 0$ and thus large flavor-violating coupling in the $\tau-e$ sector, $C_{\tau e}\approx2/3$. All non-zero axion-lepton couplings of the models E1 and E2, after imposing $C_e= C_{\mu e}= 0$, are summarized in Table~\ref{2HDM:leptonCharges}. While both models predict the same value of $C_{\tau e}=2/3$ and $C_\tau=-1/3$, the axion-muon coupling  $C_\mu$ can be either $-1/3$ or $2/3$ depending on the model. 

\begin{figure}[t]
\begin{center}
 \includegraphics[width=0.9\textwidth]{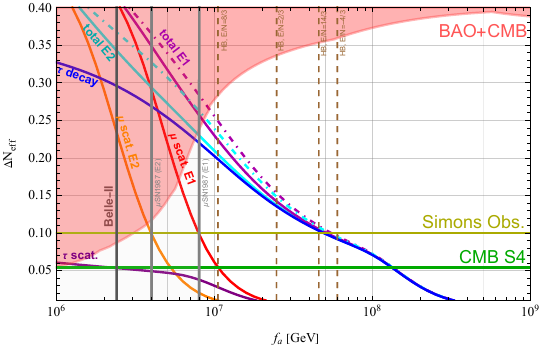}
 \caption{Additional effective number of neutrinos $\dNeff$ as a function of $f_a$ for astrophobic 2HDMs. Shown is total prediction for E1 (dark magenta) and E2 (light blue) models, without axion-kaon scatterings (solid lines) and with axion-kaon scatterings for $C_s=-2/3$ for $z=0.49$ (dash-dotted lines). The difference in $\dNeff$ between E1 and E2 models is entirely due to the different axion coupling to muons, which for E1 (E2) is $C_\mu=-2/3\ (1/3)$. Also shown are the predictions separately from $\tau \to e$ decays (blue), $\tau$-scatterings (violet) and $\mu$-scatterings, which differ between E1 (red) and E2 (orange). Vertical lines exclude the region to their left, and show constraints on the muon coupling from SN1987A~\cite{Croon:2020lrf} (grey), the Belle-II limit on $C_{\tau e}$~\cite{Belle-II:2022heu} (cyan) and   the bound on $C_\gamma$ from Horizontal Branch stars (HB)~\cite{Ayala:2014pea} (brown, dashed). 
 }
\label{fig:dNeff_2HDM}
\end{center}
\end{figure}

An astrophobic model is then obtained by combining any model in Table~\ref{2HDM:quarkCharges} with any model in Table~\ref{2HDM:leptonCharges}, so in total there are eight such models, which we denote as e.g. Q1E1 with obvious notation. In Tables~\ref{2HDM:quarkCharges} and~\ref{2HDM:leptonCharges} we also list the contributions from each sector to the electromagnetic anomaly coefficient  for each model, denoted by $E_Q/N$ and $E_L/N$. The total contribution to the electromagnetic anomaly coefficient is given by the sum $E/N=E_Q/N+E_L/N$, for example model Q1E1 predicts $E/N=8/3$. 
Simultaneous suppression of nucleon and electron couplings in the eight astrophobic  2HDMs then essentially fixes the flavor-violating axion coupling $C_{\tau e} \approx 2/3$,  which dominates the contribution to $\dNeff$ in all these models. We show the resulting prediction for $\dNeff$ as a function of $f_a$ in Fig. \ref{fig:dNeff_2HDM}, where we also display  the separate contributions to $\dNeff$ from flavor-violating $\tau$-decays and flavor-diagonal $\tau$- and $\mu$-scattering. Note that these contributions differ only between the two possible choices for the lepton sector, E1 and E2, giving the total lepton contribution to $\dNeff$ that is denoted by solid lines (``total E1/E2"). There is also a small contribution from kaon scattering,  giving the total contribution to $\dNeff$ denoted by dash-dotted lines.  Also shown is the Belle-II limit on flavor-violating $\tau e$ couplings, which requires $f_a \gtrsim 2 \times 10^6 \GeV$. 

Although constraints on axion couplings to nucleons and electrons are avoided in these astrophobic models by construction, they are also subject to constraints on axion couplings to muons (from SN1987A) and photons (from HB stars\footnote{Note that the CAST bound~\cite{CAST:2017uph} is significantly weakened below $f_a \sim 10^8 \GeV$.}), shown as vertical lines in Fig.~\ref{fig:dNeff_2HDM}. The SN1987A bound only depends on the chosen lepton scenario, and allows for $f_a$ as low as about $8\times10^6\GeV$ ($4\times 10^6\GeV$) in  E1 (E2). The constraint from HB stars is more stringent, and is controlled by the photon coupling, which can take only four different values (among the eight models) determined by  $E/N\in\{14/3, 8/3, 2/3,-4/3 \}$. In  Fig.~\ref{fig:dNeff_2HDM} we show the resulting bound on $f_a$ for these four representative scenarios as vertical dashed brown lines, which varies from about $10^7\GeV$ (e.g. Q1E1) to $6\times 10^7\GeV$ (Q4E1). These limits are even more constraining than future Belle II searches for $\tau \to e a$ that will probe up to $f_a \sim 8 \times 10^6 \GeV$.  This implies that current constraints on $f_a$ allow for $\dNeff$ as large as 0.20 (0.22) for E2 (E1) models with $E/N=8/3$, only considering leptonic production. However, for $f_a =\mathcal{O}(10^7) \GeV$ contributions to $\dNeff$ from axion-kaon scattering cannot be neglected, which arise from non-zero axion couplings to strange quarks. The maximal contribution from such scatterings is obtained for $C_s=2/3$, which increases the total prediction for $\dNeff$  for E2 (E1) models with $E/N=8/3$ up to about 0.23 (0.25).
For $f_a\gtrsim5\times10^7 \GeV$ all models give essentially the same prediction for $\dNeff$, because only $\tau$-decays are relevant for such large $f_a$.  CMB-S4 will probe the parameter space up to $f_a \simeq 10^8 \GeV$, which translates to $m_a \simeq0.06$ eV, where the axion can indeed be considered relativistic at recombination to good approximation.

\begin{figure}[t]
	\begin{center}
		\includegraphics[width=0.75\textwidth]{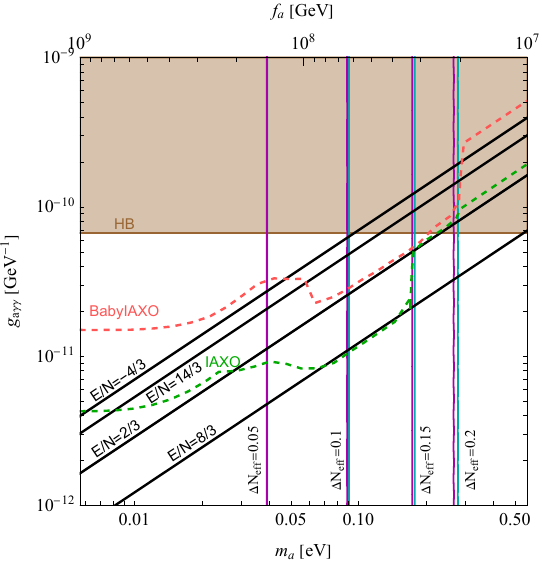}
		\caption{Predictions for the axion-photon coupling $g_{a\gamma\gamma}$ for the astrophobic models listed in Tables~\ref{2HDM:quarkCharges} and \ref{2HDM:leptonCharges}, characterised by the four indicated values of $E/N$. Magenta and cyan vertical lines indicate the total prediction for $\dNeff$ (including axion-kaon scattering for $z=0.49$)  for E1 and E2 models, respectively. We denote in brown the region excluded by HB stars constraints, while the expected sensitivities of BabyIAXO and IAXO \cite{IAXO:2020wwp} will extend this region down to the contours plotted in dashed light red and green, respectively. 
		 }
		\label{fig:gagg_vs_f}
	\end{center}
\end{figure}

It is interesting to compare these expectations with complementary probes by future helioscopes such as BabyIAXO and IAXO. In Fig.~\ref{fig:gagg_vs_f} we show these prospects in the usual $(m_a, g_{a \gamma \gamma})$ plane, for the relevant $f_a$-window between $10^7 \GeV$ and $10^9 \GeV$. The four scenarios representing the predictions of astrophobic 2HDMs are denoted by black lines, and we show in brown the excluded region from HB star cooling comstraints, the future projections for BabyIAXO (red) and IAXO (green) lines, and the contour lines for $\Delta N_{\rm eff}$ in magenta (E1) and cyan (E2).  BabyIAXO will constrain only models with $E/N = 14/3$ or $E/N = -4/3$, up to  $f_a \simeq 10^8 \GeV$, which roughly matches the reach expected from CMB-S4. IAXO instead will probe the same models down to $f_a \simeq 10^9 \GeV$, way below the $N_{\rm eff}$ sensitivity.  The other two scenarios with $E/N = 2/3$ and $E/N = 8/3$ have smaller photon couplings, and will be complementarily probed by IAXO and CMB-S4, reaching scales of $f_a \simeq 2 \times 10^8 \GeV$ and $f_a \simeq 8 \times 10^7 \GeV$, respectively. Note, however, that for $m_a\gtrsim0.2 \eV$ IAXO substantially loses its sensitivity to models with $E/N = 8/3$, and a small range of $m_a$ up to about $0.5$~eV will only be probed by future CMB surveys. Interestingly, also assuming that axions make up all dark matter in the Universe (which is possible even for $f_a\approx10^7\GeV$ in various cosmological scenarios with non-trivial evolution of the axion field~\cite{Turner:1985si,Lyth:1991ub,Hiramatsu:2012sc,Kawasaki:2014sqa,Co:2017mop,Baratella:2018pxi,Co:2018mho,Harigaya:2019qnl,Co:2019jts,Redi:2022llj,Harigaya:2022pjd,Niu:2023khv}), the region inaccessible to IAXO will be covered by JWST~\cite{Roy:2023omw} (see also Refs.~\cite{Bessho:2022yyu,Janish:2023kvi,Yin:2024lla}).

\begin{figure}[t]
	\begin{center}
		\includegraphics[width=0.49\textwidth]{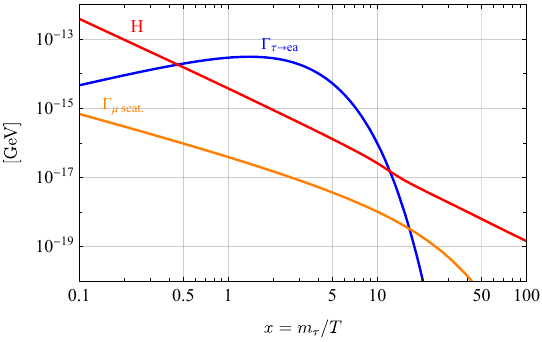}
		\includegraphics[width=0.49\textwidth]{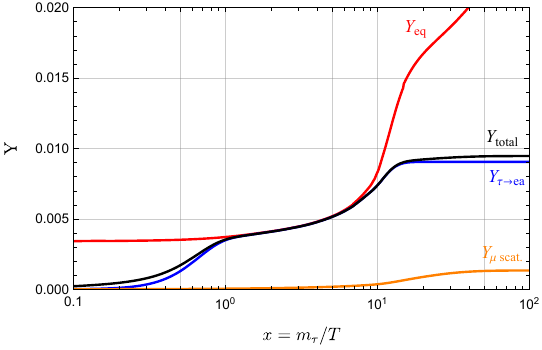}
		\caption{\textit{Left panel:} Axion production rates from $\mu$-scatterings (orange) and $\tau$-decays (blue) as a function of $x=m_\tau/T$ compared to the Hubble rate (red) for E1 models ($C_\mu=-2/3$) at $f_a=2\times 10^7\GeV$. \textit{Right panel:} Axion yield  obtained by numerically solving the Boltzmann equation  taking into account scattering and decays simultaneously (black) and separately (orange/green), compared to the equilibrium yield (red). 
		}
		\label{subsqnt_freeze-in}
	\end{center}
\end{figure}  

We close this section with a more detailed discussion of the various contributions to $\dNeff$ in Fig.~\ref{fig:dNeff_2HDM}. Even though in 2HDM $\tau$-decays dominate thermal axion production, the total $\dNeff$ exceeds the value predicted when considering only  such decays below $f_a=3\times 10^7\GeV$. Indeed there is a sub-leading contribution to axion production from $\mu$-scatterings, which freeze-in axions after inverse $\tau$-decays have frozen out.
Axion production from $\tau$-decays and $\mu$-scatterings take place at different temperatures, affecting the abundance in a non-trivial way. In the left panel of Fig. \ref{subsqnt_freeze-in} we show the leptonic rates in E1 models compared to the Hubble rate for a representative value of $f_a=2\times 10^7\GeV$. As anticipated, the $\tau$-decay rate is large enough to keep axions in thermal equilibrium for $1 \lesssim x \lesssim 10$. The $\mu$-scatterings instead are never effective enough to bring axions into thermal equilibrium, but still produce axions via freeze-in. 
On the right panel of Fig. \ref{subsqnt_freeze-in} we show the evolution of the axion co-moving number density for the same benchmark value of $f_a=2\times 10^7\GeV$. The rapid change in $Y_\textrm{eq}$ is a result of the sudden changes in the number of relativistic degrees of freedom around the QCD phase transition at $x_\text{QCDPT}\approx11$. As expected, the equilibrium yield is reached due to (inverse) $\tau$-decays and $\dNeff$ is determined by the freeze-out temperature $x\sim 15$. However, for $x>15$ $\mu$-scatterings become the main production channel, adding on top of the axion yield produced from $\tau$-decays. Note however that the subsequent freeze-in is less effective than without $\tau e$-couplings, because the collision term is proportional to $Y^\text{eq}-Y$, and thus reduced as compared to the pure $\mu$-scattering scenario due to the non-zero initial axion abundance. Hence, while axion freeze-in production solely from $\mu$-scattering gives $\Delta Y\sim 1.5\times 10^{-3}$, the same production subsequent to freeze-out production from $\tau$-decays gives only $\Delta Y\sim 0.5\times 10^{-3}$.

\section{ $\dNeff$ in Astrophobic 3HDMs}
\label{sec:3HDM}

In the previous section we have discussed non-universal DFSZ models with two Higgs doublets, which do not allow for suppressed  axion couplings to muons or photons, in addition to nucleons and electrons. This is the reason why in those models $f_a$ cannot be smaller than few times  $10^7 \GeV$. As we will show now, this restriction can be avoided on the price of adding a third Higgs doublet, which admits $f_a\lesssim 10^7$~GeV without violating any astrophysical constraints. Most of these models can be probed by improved measurements of $\dNeff$, and require only a mild tuning to  simultaneously suppress axion couplings to nucleons, electron and photons, in contrast to 2HDMs. 

There are many 3HDMs giving astrophobic axions, and we refer for a detailed discussion to Appendix~\ref{models}. In this section we classify them according to the suppressed couplings relevant for avoiding astrophysical constraints: i) $|C_\mu| \ll 1$ ii) $|C_\gamma| \ll 1$, and iii) $|C_\mu|, |C_\gamma| \ll 1$ with LFV.
The case of $|C_\mu|, |C_\gamma| \ll 1$ without LFV will be discussed in Sec.~\ref{sec:natural} in the context of the naturally astrophobic axion model. For each model we will compute the axion contribution to $\dNeff$ as a function of $f_a$. All models are obtained by combining a quark sector model in Table~\ref{Qmodels3} with a lepton sector model in Table~\ref{Lmodels3}, and require a single small PQ charge $X_0 \ll 1$ in order to be astrophobic. This can be achieved by choosing appropriate Higgs vacuum angles,  $X_0 = (1- 3 c_1^2) c_2^2$ (cf. Eq.~\eqref{X0} and the discussion below), and the possible degree of suppression is only bounded by perturbativity of Yukawa couplings. Predictions for axion couplings in lepton sectors of astrophobic 3HDMs after imposing conditions for astrophobia are presented in Table~\ref{3HDM:leptonCharges}. Note that E1 and E2 models in the context of 3HDMs give different predictions for axion couplings to leptons than E1 and E2 in the context of 2HDMs (cf. Table~\ref{2HDM:leptonCharges}). The predictions for axion couplings to quarks are given in Table~\ref{2HDM:quarkCharges} and for Q1-Q4  are the same within 2HDMs and 3HDMs. There is a single new model in the context of 3HDMs, dubbed Q5.

\begin{table}[t]
	\centering
	\begin{tabular}{|c||c||c|c||c|}
		\hline
		Model & $E_L/N$   & $C_\mu$ & $C_\tau$& $C_{\tau \mu}$    \\ \hline
		E1 & -4/3  & 0 & -2/3 &   $0$ \\ \hline 
		E2 &  2/3 & 0 & 1/3 &  $0$ \\ \hline 
		E3 &  0 & 0 & 0 &  $0$ \\ \hline  
		E4 &  -8/3 & -2/3 & -2/3 &  $0$ \\ \hline 
		E5 &  4/3 & 1/3 & 1/3 &  $0$ \\ \hline  \hline
		E6 &  -2/3 & 0 & 1/3 &  $2/3$ \\ 
		\hline 
	\end{tabular}
\caption{Lepton sectors in astrophobic 3HDM axion models after imposing conditions for nucleophobia,  $C_u=2/3$ and $C_d=1/3$ and electrophobia $C_e=0$. In models E1 and E2, conditions for muonphobia $C_\mu=0$ were additionally imposed. In models E1-E5, the conditions for nucleophobia, electrophobia and muonphobia (where possible) imply no LFV couplings. The E6 model  is special as all three leptons carry different PQ charges.}
\label{3HDM:leptonCharges}
\end{table}

\subsection{Models with $|C_\mu|\ll1$}\label{3HDM:suppressedMuon}
 
 \begin{figure}[t]
	\begin{center}
		\includegraphics[width=0.9\textwidth]{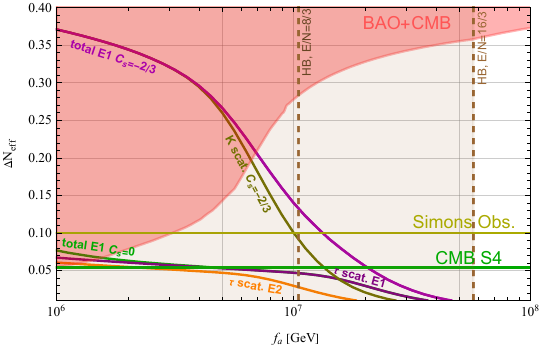}
		\caption{Additional effective number of neutrinos $\dNeff$ as a function of $f_a$ for astrophobic 3HDMs with $|C_\mu|\ll1$. These models are obtained by combining E1, E2 and E3 with any model of the quark sector (Q1-Q5), and differ in their contributions to $\dNeff$  only through the different values of $C_\tau$ and $C_s$. We only show the largest possible contribution represented by model Q3E1. The most important lower bound on $f_a$ comes from HB star cooling bounds on the axion-photon coupling~\cite{Ayala:2014pea}, which is strongest (weakest) for $E/N= 16/3 (8/3)$.
		}
		\label{fig:dNeff5.1}
	\end{center}
\end{figure}  
 
We start with a discussion of models in which the axion coupling to muons is small, $|C_\mu|\ll1$, but the axion-photon coupling is unsuppressed. In these models the PQ charges of leptons can have a $2+1$  flavor structure. There are 14 models of this type that are obtained by combining any of the 5 potentially nucleophobic models in  the quark sector, summarized in Table~\ref{Qmodels3}, with either model E1, E2 or E3 in the charged lepton sector\footnote{Model E6 instead has a 1+1+1 flavor structure and will be discussed below.}, defined in Table~\ref{Lmodels3}, and taking $X_0 \ll 1$ and $\xi^{e_L}_{33} =1$, so that there is no LFV. Models involving E3 are special because all axion-lepton couplings vanish.  The model Q5E3, in which also the axion-photon coupling is  suppressed, will be discussed separately in Section~\ref{sec:natural}.  Nucleophobia and electrophobia can thus be obtained without flavor-violation, upon making appropriate choices for quark flavor rotations. Therefore the only leptonic contribution to $\dNeff$ comes from the axion-tau coupling, which is fixed to be $C_\tau=-2/3$ in E1 and $C_\tau = 1/3$ in E2 models, respectively, and for E3 model even this contribution is absent.

The main difference between the five possible choices of quark sector models is the value of the axion-photon coupling. Taking into account all possible combinations with the three lepton sector models, this coupling is determined by seven representative values, $E/N\in\{16/3, 14/3, 10/3, 8/3, 4/3, 2/3, -2/3 \}$. These translate into a lower bound on $f_a$ from HB constraints, which varies from about $10^7$~GeV for the least constrained models with $E/N=8/3$  (Q2E3, Q3E3, Q5E2) and $E/N=4/3$ (Q2E1, Q3E1, Q4E2)  to about $6\times10^7$~GeV for the most constrained model (Q1E2) with $E/N=16/3$.

Since electron, muon, and LFV couplings are suppressed, the only leptonic contribution to $\dNeff$ arises from scattering processes of $\tau$-leptons, differing only between models E1 ($C_\tau = -2/3$), E2 ($C_\tau = 1/3$) and E3 ($C_\tau = 0$), and the resulting predictions are shown in Fig.~\ref{fig:dNeff5.1}. Taking into account the lower bounds on $f_a$ from HB constraints as discussed above, this contribution is always below the sensitivity of  CMB-S4.
However, because there are no LFV couplings, the largest contribution to  $\dNeff$ in these models are actually due to kaon scattering, which is controlled by the value of the axion couplings to $s$-quarks. The value of this  coupling can take just three different values, $C_s = 1/3$ or $C_s = -2/3$ in models Q1-Q4, while it vanishes in Q5. In Fig.~\ref{fig:dNeff5.1} we show separately the predictions for  $\dNeff$ from tau and kaon scattering for two selected models in order to avoid clutter, which correspond to the smallest ($C_s = 0$ in Q5E1) and largest contributions from kaon scattering ($C_s = -2/3$ in e.g. Q3E1). The maximal contribution  to $\dNeff$ in these models, after taking into account the constraints on the axion-photon coupling, is 0.13 and within the reach of the Simons Observatory. However, the potential for improvement of the lower bound on $f_a$ by future CMB experiments is quite limited and CMB-S4 will improve it up to $2\times10^7\GeV$ (for models with maximal $\dNeff$), which is better than the astrophysical bound only for models with $E/N=8/3$~or~$4/3$.
The other models will be probed more efficiently by IAXO or JWST than by CMB observations.

\subsection{Models with $|C_\gamma|\ll1$}

\begin{figure}[t]
	\begin{center}
		\includegraphics[width=0.9\textwidth]{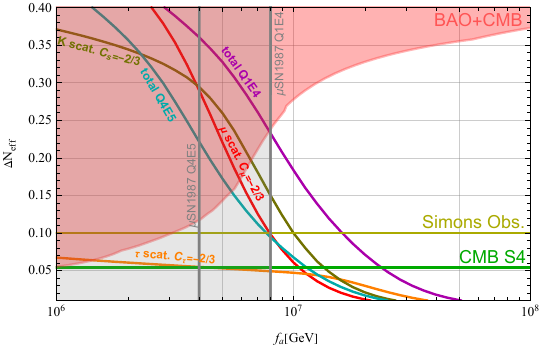}
		\caption{Additional effective number of neutrinos $\dNeff$ as a function of $f_a$ for astrophobic 3HDMs with $|C_\gamma|\ll1$, which is realized in Q1E4 and Q4E5 models. The vertical grey line denotes the limit on the muon coupling from SN1987A \cite{Croon:2020lrf}, while red, brown and orange lines denote the contributions from scattering off muons, kaons and taus, respectively.}
		\label{fig:dNeff5.2}
	\end{center}
\end{figure} 

Assuming a $2+1$ flavor structure of the (non-universal) PQ lepton charges, it is also possible to suppress the axion-photon coupling in 3HDMs. Indeed there are two models with $E/N=2$ (Q1E4 and Q4E5), which allow for $f_a\sim \mathcal{O}(10^6) \GeV$ without being in conflict with HB star cooling constraints. However, in these models axion-electron and axion-muon couplings cannot be suppressed simultaneously, as electrophobia implies $C_\mu=C_\tau=-2/3$ (1/3) in Q1E4 (Q4E5) model, which in turn requires  $f_a\gtrsim8(4)\times10^6$~GeV from the SN1987A constraint on the axion-muon coupling. There is no LFV in these models, so that the only sizable contribution to $\dNeff$ comes from axion-kaon, axion-muon, and axion-tau scatterings, and we show these contributions and the total prediction for $\dNeff$ in Fig.~\ref{fig:dNeff5.2}. The predicted value of $\dNeff$ in the Q1E4 model is larger than in the Q4E5 model, since the absolute values of axion couplings to muon, tau, and strange are twice as large in Q1E4 as compared to  Q4E5. We see that the combined effect of these three channels leads to $\dNeff$ as large as 0.23 without violating the SN1987A constraint, which exceeds the maximal value that can be obtained in the models with $|C_\mu| \ll 1$ as discussed above, cf.~Fig.~\ref{fig:dNeff5.1}. Interestingly, the current cosmological data set lower bounds on $f_a$ in these models that are comparable to those from astrophysics. In Q4E5 the lower bound on $f_a$ from thermal axion production is about $5\times10^6$~GeV so even slightly stronger than the SN1987A constraint. This is because for such small $f_a$ the axion mass is around 1~eV, so the axion is not relativistic around recombination,  which strenghens the upper bound on $\dNeff$. Due to relatively weak astrophysical constraints there are good prospects for testing both models in future CMB surveys.  Model Q1E4 can be within the reach  of the Simons Observatory (CMB-S4) for values of $f_a$ up to $10^7$ ($2\times10^7$)~GeV, while the reach for $f_a$ in Q4E5 is weaker by about a factor two. Both models cannot be probed by IAXO due to the suppressed axion-photon coupling. 

On the other hand, these models may still be probed by JWST, which for $f_a<10^7$ GeV will be sensitive to the axion-photon couplings $g_{a\gamma\gamma}\equiv\alpha(2 \pi f_a)^{-1}C_{\gamma}$ as small as $\mathcal{O}(10^{-11})$~\cite{Roy:2023omw}. However, whether JWST can really observe such axions depends on the contribution to $C_\gamma$ from axion-pion mixing, which has rather large uncertainties. For the central value of this contribution $g_{a\gamma\gamma}\approx10^{-11}$ for $f_a=8\times10^6$~GeV, which is on the verge of the JWST sensitivity, but $g_{a\gamma\gamma}$ could be a factor of two larger when taking into account theoretical errors. Still, for $E/N=2$ theoretical uncertainties allow for vanishing $g_{a\gamma\gamma}$ so definite conclusions cannot be made until the theory prediction has improved.

\subsection{Models with $|C_\mu|\ll1$ and $|C_\gamma|\ll1$ and LFV}

We finally discuss {\it proper} astrophobic models, where on top of nucleophobia and electrophobia also axion couplings to muons and photons are suppressed, so that {\it all} stellar cooling constraints are weakened. In such models, the dominant lower bound on $f_a$ originates from the usual SN1987A constraint on axion-nucleon couplings, which are induced by higher-order corrections. Taking into account these corrections, the resulting lower bound on $f_a$ can be as weak as $10^6 \GeV$, although the exact numerical value depends on the details of the axion model and is also sensitive to the uncertainties in the lattice determination of the ratio $m_u/m_d$~\cite{Badziak:2023fsc}.

As discussed in Appendix~\ref{models}, astrophobic models with $|C_\mu|\ll1$ and $|C_\gamma|\ll1$ can be constructed within the framework of DFSZ-like models with 3 Higgs doublets. There are three such models, depending on the PQ charge structure of charged leptons. Either the charges are universal, so that there is no LFV, or they are different for each generation, i.e., have a 1+1+1 flavor structure, indicating possibly large LFV. Here we focus first on the two proper astrophobic axion models with LFV, which have the best prospects to be probed by CMB observations in the near future, since flavor-violating $\tau$-decays give the dominant contribution to $\dNeff$, in the absence of the axion-pion coupling. We will discuss the model with universal lepton charges (Q5E3) in the next section.

\begin{figure}[t]
	\begin{center}
		\includegraphics[width=0.9\textwidth]{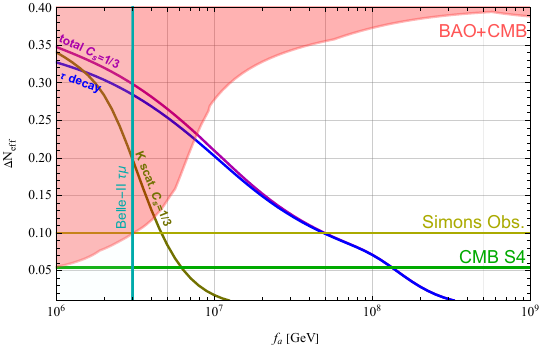}
		\caption{Additional effective number of neutrinos $\dNeff$ as a function of $f_a$ for proper astrophobic 3HDM with LFV, realized by the Q2E6 and Q3E6 models. The plot shows the prediction in the model with the smallest kaon contribution ($C_s = 1/3$) and $z=0.49$. The maximal contribution for $C_s = -2/3$ to $\dNeff$  is slightly larger, but the resulting current bound on $f_a$ from cosmology is very similar. The vertical cyan line marks the lower bound on $f_a$ from searches for flavor-violating $\tau \to\mu a$ decays at Belle-II \cite{Belle-II:2022heu}.}
		\label{fig:3HDM_LFV}
	\end{center}
\end{figure}

The two models are obtained by combining E6 in Table~\ref{Lmodels3} with models Q2 or Q3 in Table~\ref{Qmodels3} (dubbed Q2E6 and Q3E6, respectively), giving $E/N = 2$ and thus a suppressed axion coupling to photons. Proper astrophobia is obtained for $X_0 \ll 1$, giving for both models $C_e = C_\mu =0$, $C_\tau=1/3$, and  the LFV coupling $C_{\tau\mu}=2/3$. The largest contribution to $\dNeff$ is due to axion production from $\tau$-decays, but a sub-leading contribution is due to kaon scattering, which depends on the value of $C_s$ that is controlled by quark flavor rotations, and can vary between $-2/3$ and $1/3$. In Fig.~\ref{fig:3HDM_LFV} we show the minimal total prediction for $C_s = 1/3$, as well as the contributions from kaon scattering and $\tau$-decays alone.  
In contrast to the models discussed above, the constraint on $f_a$ provided by cosmology, i.e., CMB and BAO data,  is much stronger than limits from astrophysics or direct searches for LFV decays. 
Interestingly, despite the suppression of the axion-muon and axion-photon couplings, this only excludes values $f_a\lesssim8\times10^6 \GeV$. This is mainly due to the rather large production rate of thermal axions from $\tau\to \mu a$ decays, as well as the fact that for $f_a\lesssim10^7$~GeV axions act as a warm dark matter and the upper bound on $\dNeff$ rapidly tightens when decreasing $f_a$.

These models can be complementary probed by searches for $\tau\to \mu a$ at Belle-II~\cite{Belle-II:2022heu}. While  current bounds are not competitive with the constraints from cosmology, future runs of Belle-II are expected to probe $f_a$ up to the level of $10^7$~GeV, which corresponds to $\dNeff \simeq 0.2$ for the models under consideration, and thus allow for complementarity to 
future CMB searches. Similarly to the models Q1E4 and Q4E5 with $|C_\gamma|\ll1$ and unsuppressed axion-muon couplings discussed above (cf.~Fig.~\ref{fig:dNeff5.2}), the JWST sensitivity strongly depends on the uncertainty of the theoretical prediction for the axion-photon coupling.

\section{ $\dNeff$ from the Naturally Astrophobic QCD Axion}
\label{sec:natural}

In this section, we discuss the naturally astrophobic axion model~\cite{Badziak:2023fsc}. In this model, the SM Higgs is a nearly PQ charge eigenstate with a vanishing PQ charge, and the axion coupling to  SM fermions is entirely controlled by their PQ charge assignments, so that no tuning of the parameters of the theory is required to achieve astrophobia.

In particular, a proper astrophobic model without LFV can be achieved by assigning the PQ charges of 2, 1, 0, and 0 to $u$, $d$, $e$, and $\mu$, respectively, and assuming that the QCD and electromagnetic anomaly comes only from $u$ and $d$. In the minimal scenario, the PQ charges of other SM fermions are zero. In the UV completion by 3HDM, this can be obtained by combining the flavor-universal model E3 in the charged lepton sector (see Table~\ref{Lmodels3}) with model Q5 in the quark sector, (see Table~\ref{Qmodels3}), upon taking $X_0 \ll 1$ and $\xi^{u_R}_{11} = \xi^{d_R}_{11} = 1$. This model, dubbed Q5E3, has a very interesting feature that the axion couples exclusively to $u$- and $d$-quarks, so that $X_0 \ll 1$ can be realized by strongly suppressing the vevs of the two non-SM Higgs fields consistent with perturbative unitarity, as these only give rise to up- and down-quark masses. This is in contrast to all other 3HDM models, where perturbativity of Yukawa couplings prevents this possibility, and instead require some (mild) tuning, see e.g. Ref.~\cite{Bjorkeroth:2019jtx}.  This model was proposed in Ref.~\cite{Badziak:2023fsc}, and it was shown that it can be UV-completed not only within the 3HDM scenarios but also by adding new vector-like quarks. 

\begin{figure}[t]
	\begin{center}
		\includegraphics[width=0.85\textwidth]{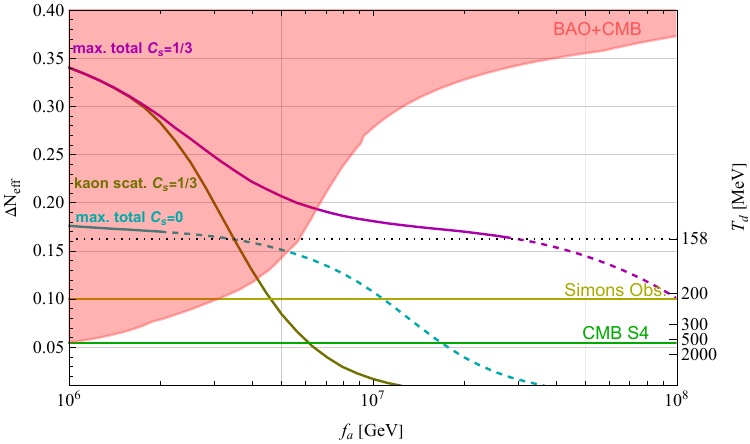}
		\caption{Additional effective number of neutrinos $\dNeff$ as a function of $f_a$ for the naturally astrophobic axion, realized e.g.  by the E3Q5 model. Dotted lines indicate uncertainties due to the non-perturbative freeze-in. The axion-kaon scattering has been computed for $z=0.49$ for which $f_a$ down to few $\times 10^6$ GeV is consistent with astrophysical constraints.  Axion decoupling temperatures $T_d$ assuming prior equilibrium above the QCDPT are shown on right vertical axis. 
		}
		\label{fig:dNeff_natural}
	\end{center}
\end{figure} 

Both astrophysical constraints on $f_a$ and the predictions for $\dNeff$ somewhat depend on the particular UV completion. We show the prediction of $\dNeff$ for the naturally astrophobic axion in various scenarios in Fig.~\ref{fig:dNeff_natural}.
In the minimal model only up- and down-quarks couple to the axion and the corresponding couplings are $C_u=2/3$ and $C_d=1/3$.
In the minimal model the astrophysical constraints on the axion-nucleon couplings allow for $f_a$ as small as $10^6\GeV$ if $z\approx0.49$~\cite{Badziak:2023fsc}. In this scenario axion couplings to pions, leptons and kaons are strongly suppressed, so the axion cannot be in thermal equilibrium below the QCD phase transition. Therefore, non-negligible contribution to $\dNeff$ may only arise in the deconfined phase when the axion may be kept in thermal equilibrium by scattering with up- and down-quarks. The production rates of axions in these scattering processes with quarks are proportional to temperature, so the dominant production of axions occurs at low temperatures (since the Hubble scale scales as $T^2$). Hence, we expect that axions will be mostly produced for temperatures not far above the QCD phase transition. Unfortunately, the scattering rates for these processes cannot be reliably computed for such temperatures using perturbative methods~\cite{Notari:2022ffe} (see also Refs.~\cite{Graf:2010tv,Salvio:2013iaa}). On the other hand, we know that astrophobic axions are not thermally produced in the regime when chiral perturbation theory correctly describes axion interactions. Thus, it is possible to estimate the maximal contribution to $\dNeff$ that occurs if the axion decouples from SM plasma for temperatures around 150 MeV.  However, if the production rate of axions is smooth it may well be that axions decouple at some higher temperature. For this reason in Fig.~\ref{fig:dNeff_natural} we also show the relation between $\dNeff$ and axion decoupling temperature using an instantaneous decoupling approximation. We see that if the strong interactions keep the axion in thermal equilibrium for temperatures down to the QCD phase transition, $\dNeff$ can be as large as about 0.15, while if it decouples around 1 GeV, $\dNeff$ is around 0.05. 

In Fig.~\ref{fig:dNeff_natural} we also show the maximal value of $\dNeff$ obtained by estimating the non-perturbative effects on axion production using the leading order up- and down-quark scattering rates above the QCD phase transition~\cite{DEramo:2021lgb}, setting the strong coupling $g_s$ to $4\pi$. We see that in the minimal scenario (corresponding to $C_s=0$ in the plot), such scatterings are not able to thermalize axion unless $f_a$ is below few $\times 10^6\GeV$. Nevertheless, freeze-in axion production via such scatterings may give non-negligible contribution to $\dNeff$ resulting in the current lower bound from Planck around $5\times10^6\GeV$, which should be considered as an aggressive limit for the minimal model.

Given the uncertainty in the prediction of $\dNeff$ one cannot use future CMB data to set robust bounds on the naturally astrophobic axion scenario. On the other hand, in case some deviation from $\Lambda$CDM model is found it may still be explained by the minimal model with $f_a=\mathcal{O}(10^7)\GeV$, but to reliable extract information about the axion from such data one would require a lattice determination of the axion temperature evolution.

 The minimal model, where $C_d=1/3$ and $C_s=0$, requires additional model building (see Ref.~\cite{Badziak:2023fsc}) to naturally suppress rare kaon decays induced by flavor-violating axion couplings in the $s$-$d$ sector. Instead such decays are naturally suppressed without  model building if $C_s=C_d$.  For $C_s=1/3$, the axion contribution to $\dNeff$ is larger than in the minimal model for two reasons. First, the axion-kaon scattering is no longer suppressed and results in a lower bound on $f_a$ from Planck around $4\times10^6\GeV$, as seen from Fig.~\ref{fig:dNeff_natural}. Second, axion production from strange-quark scattering is expected to be bigger than that from up- and down-quark scattering due to much larger strange-quark mass. Using the same estimate for the non-perturbative contribution as above, we found that strange-quark scattering can keep the axion in thermal equilibrium for temperatures down to the QCD phase transition for $f_a$ exceeding even $10^7\GeV$. This results in a lower bound on $f_a$ from Planck about  $7\times10^6\GeV$, as seen in Fig.~\ref{fig:dNeff_natural}. 
  
 We have also checked that the lower bound of $f_a\gtrsim 7\times10^6\GeV$ from Planck is independent of the choice for estimating the non-perturbative axion production. For example, the bound stays the same if one takes $g_s=\sqrt{4\pi}$ instead of $4\pi$. Also the estimate proposed in Ref.~\cite{Notari:2022ffe} with the thermally averaged rate parameterized as $\bar{\Gamma}=\kappa T^3/f^2$ for temperatures between $2\GeV$ and the QCD phase transition with $\kappa=0.1$ or $0.01$ gives the same lower bound on $f_a$. In all these estimates the axion is thermalized down to the QCD phase transition for $f_a=\mathcal{O}(10^7)\GeV$, which is the reason why the exact value of the rate has no impact on the predicted value of $\dNeff$. On the other hand, the estimates for the sensitivity of future experiments are much less robust.  For example, the reach of the Simons Observatory for $f_a$ varies between $10^7$ and $10^8$~GeV if one changes $g_s$ from $\sqrt{4\pi}$ to $4\pi$ in our estimate for strange-quark scattering.

Let us also comment on the fact that for $C_s=1/3$ the astrophysical bounds on the axion-nucleon couplings can allow for $f_a$ much below $10^7\GeV$ only if axion couplings to charm and/or bottom are also $\mathcal{O}(1)$~\cite{Badziak:2023fsc}. We have checked that turning on these couplings does not affect the current lower bound on $f_a$. This is because axion production from charm- and bottom-quark scattering is Boltzmann suppressed around the QCD phase transition, and also because strange-quark scattering alone is sufficient to thermalize axions down to the QCD phase transition.

We emphasize that the naturally astrophobic axion is the model that is least constrained by the CMB, and $f_a$ much below $10^7$~GeV may be consistent with all available data. Such values of $f_a$ imply an axion mass $\mathcal{O}(1)$~eV, so it is non-relativistic at recombination. Thus, in this scenario $\dNeff$ is just a useful parametrization of the energy density of thermal axions and axions act like warm dark matter rather than constituting extra relativistic degree of freedom, which is yet another feature that could help to distinguish the naturally astrophobic axion from other axion models using future CMB data.           

Our results also have implications to minimal axiogenesis~\cite{Co:2019wyp}. 
In the minimal axiogenesis scenario, the PQ symmetry-breaking field rotates in field space in the early universe, which corresponds to a non-zero PQ charge. The PQ charge is transferred to baryon asymmetry via axion-SM couplings and electroweak sphaleron processes. For $f_a \gg 10^7$ GeV, however, the produced baryon asymmetry is smaller than the observed one after imposing the constraint from overproduction of axion dark matter by the kinetic misalignment mechanism~\cite{Co:2019jts}, unless some of the axion-SM  couplings are much larger than the $1/f_a$-suppressed one. For $f_a \lesssim 10^7$ GeV, on the other hand, the observed baryon asymmetry can be explained without introducing large couplings. Even with the maximal possible scattering rate of the axion with $u$- and $d$-quarks or $C_s=1/3$, the current constraints from the CMB and BAO can be satisfied. If the scattering rates with quarks become as large as those with $g_s \sim 4\pi$ before the QCD phase transition, CMB-S4 can probe the parameter space of the successful minimal axiogenesis without large couplings.

Even though axion-photon coupling in this scenario is smaller by at least one order of magnitude than in models with $E/N\neq2$, it may be still possible for JWST to discover a DM axion with mass of $\mathcal{O}(1) \, {\rm eV}$. Thus, using complementarity of future CMB and JWST observations, it will be viable to pin down axions that are responsible both for DM and baryogenesis.

\section{Conclusions}
\label{sec:concl}

We have investigated thermal production of QCD axions in so-called astrophobic models in which astrophysical constraints on the axion decay constant are relaxed. A model-independent feature of such models is that the axion-pion coupling is so small that the impact of axion-pion scattering on the production of thermal axions is negligible. However, we found that a large abundance of hot axions, parameterised by the extra effective number of neutrinos $\dNeff$, can be sizable in the presence of other axion couplings, such as couplings to muons, strange-quarks or LFV couplings leading to axion production from $\tau\to \ell a$ decays, see Fig.~\ref{fig:dNeff_MI}. 

The simplest astrophobic axion models are generalized DFSZ models with two Higgs doublets with flavor non-universal PQ charges~\cite{DiLuzio:2017ogq}. In this class of models, suppression of axion couplings to nucleons and electrons fixes the size of LFV couplings, which directly control $\tau\to e a$ decays and give a sharp prediction for $\dNeff$ as a function of $f_a$. However, axion-photon couplings are unsuppressed and a lower bound on $f_a$ arises from  HB star cooling constraints, which varies between $(1 \div 6) \times 10^7 \GeV$, depending on the model. This in turn limits the predicted $\dNeff$ to be below about $(0.1 \div 0.25)$ (see Fig.~\ref{fig:dNeff_2HDM}), which are within the reach of the Simons Observatory. This is in sharp contrast to the predictions of the minimal DFSZ and KSVZ axion models, where $\dNeff$ is at most $0.03$ in the region of $f_a$ consistent with astrophysical constraints, and it is difficult to observe such small $\dNeff$ in the foreseeable future.  The range of $f_a$ leading to large $\dNeff$ in astrophobic 2HDM will be also probed by axion helioscopes. Two of the models are within the reach of BabyIAXO, while the remaining ones can only be probed by IAXO.  A small range of $f_a \lesssim  3 \times 10^7 \GeV$ will not be covered by IAXO, but it is within reach of JWST, if one assumes that axions make up all dark matter in the Universe.  

We have also constructed generalized DFSZ models with three Higgs doublets. In such models there is more flexibility in the structure of axion couplings as compared to 2HDMs. In particular, it is possible to suppress nucleon and electron couplings without additional tuning. In such models $\dNeff$ is typically much smaller than in 2HDMs, because flavor-conserving axion-lepton scatterings are not as efficient in producing thermal axions. However, we found that in such models axion-kaon scatterings may still lead to large $\dNeff$. This effect is particularly important in models with axion-photon couplings small enough to allow for $f_a$ down to $10^7$~GeV, and the resulting $\dNeff$ can even be above 0.1, within the reach of the Simons Observatory (see~Fig.~\ref{fig:dNeff5.1}).

In order to allow for $f_a$ below $10^7$~GeV consistent with astrophysical constraints, it is necessary to not only arrange for $E/N=2$, which accidentally relaxes the lower bound on $f_a$ from HB stars to about $10^6$~GeV, but also to suppress the axion-muon coupling in order to avoid constraints from SN1987A. In such models, $\dNeff$ sets a lower bound on $f_a$ much stronger than the astrophysical constraints, if efficient axion production from flavor-violating $\tau$-decays or axion-kaon scattering is possible.
We constructed such models and found a lower bound on $f_a$ given by $8\times10^6$~GeV (see~Fig.~\ref{fig:3HDM_LFV}).
  
We also have investigated the recently proposed naturally astrophobic QCD axion model. We found that in the minimal model where the axion couples to the SM only via the up and down quarks, the axion cannot be kept in thermal equilibrium with the SM plasma for temperatures below the QCD phase transition, due to strongly suppressed axion couplings to pions, kaons, and leptons. However, axions may still be produced thermally before the QCD phase transition in scatterings with free quarks. The production rates for such processes cannot be reliably computed using perturbative techniques, and the maximal value of $\dNeff$ can only be estimated. If the axion decouples exactly at the QCD phase transition, current cosmological constraints lead to a lower bound  $f_a\gtrsim6\times10^6$~GeV, which can be even further weakened using more conservative assumptions about the axion decoupling temperature. Thus, in the naturally astrophobic axion model both astrophysical and cosmological constraints allow minimal axiogenesis to explain both the observed DM and the baryon abundance. Future CMB surveys may probe the parameter space relevant for this scenario, where the axion mass can be as large as 1~eV. The axion-photon coupling may be large enough to lead to a discovery of axion DM  by JWST, and  future CMB data could be used to cross-check such an interpretation.

\section*{Acknowledgments}
We would like to thank Luca di Luzio, Ricardo Ferreira, Maxim Laletin, Fabrizio Rompineve and Giovanni Villadoro for useful discussions and correspondence. M.B. and R.Z. thank
the Galileo Galilei Institute for Theoretical Physics and INFN for hospitality and partial support during the completion of this work. This work was partially supported by the National Science Centre, Poland, under research
grant no. 2020/38/E/ST2/00243 (M.B. and M.L.), has received support from the European Union's Horizon 2020 research and innovation programme under the Marie Sk{\l}odowska-Curie grant agreement No 860881-HIDDeN, is partially
supported by project C3b of the DFG-funded Collaborative Research Center TRR257 ``Particle Physics Phenomenology after the Higgs Discovery" (R.Z.),
and is partially supported by Grant-in-Aid for Scientific Research from the Ministry of Education, Culture, Sports, Science, and Technology (MEXT), Japan (20H01895) and by World Premier International Research Center Initiative (WPI), MEXT, Japan (Kavli IPMU) (K.H.).
M.L. thanks the Karlsruhe Institute of Technology (KIT) for hospitality.

\appendix

\renewcommand{\theequation}{\thesection.\arabic{equation}}
\renewcommand{\thetable}{\thesection\arabic{table}}

\section{Axion Couplings to Pions and Kaons}
\label{chiPT}
\setcounter{equation}{0}
\setcounter{table}{0}

An anomalous axial rotation of light quarks $q=(u,d,s)^T$ parametrized by $Q_a$
\begin{equation}
	q\rightarrow e^{i\gamma_5 \frac{a}{2f_a} Q_a}q \, , 
\end{equation}
which for $\textrm{Tr}Q_a=1$ cancels the $aG\Tilde{G}$ term, induces an axion dependence in the quark masses \begin{equation}
	M_a=e^{ia(x)Q_a/2f_a}M_q e^{ia(x)Q_a/2f_a},
\end{equation}
where $M_q=\textrm{diag}\left( m_u,m_d,m_s\right)$. The relevant Lagrangian involving axion and $u,d,s$ quarks is given by 
\begin{equation}\label{LagrInfrared}
	{\cal L}  =\sum_{q=u,d,s}  \frac{\left( C_q - Q_a^q \right)}{2 f_a} (\partial_\mu a)\overline{q} \gamma^\mu  \gamma_5 q-\sum_q \overline{q}M_a q \, , 
\end{equation}

We can make the usual choice $Q_a=M_q^{-1}/\text{Tr}M_q^{-1}$, which avoids axion-pion mixing. Explicitly, this results in the following form of the matrix \begin{equation} \label{MatrixQa}
	Q_a= \begin{pmatrix}
		\frac{1}{1+z+w} & 0 & 0\\
		0 & \frac{z}{1+z+w} & 0\\
		0&0& \frac{w}{1+z+w}
	\end{pmatrix} \, , 
	\end{equation}
	with $z=m_u/m_d$ and $w=m_u/m_s$. On the other hand, the LO chiral lagrangian is given by \cite{DiLuzio:2020wdo,Scherer:2002tk}\begin{equation} \label{ChPTLag}
		\mathcal{L}=\frac{f_\pi^2}{4}\Bigg( \text{Tr}\big(D^\mu U^\dagger D_\mu U\big)+2 B \text{Tr} \big(U M_a^\dagger +M_a U^\dagger\big) \Bigg)+\frac{\partial_\mu a}{2 f_a}c_a \text{Tr}\bigg(\frac{i f_\pi^2}{2}\lambda_a(U\partial^\mu U^\dagger-U^\dagger \partial^\mu U) \bigg),
	\end{equation}
	where \begin{equation}
		U = \exp\left(\frac{i}{f_\pi}\begin{pmatrix}
			\pi^0+\eta/\sqrt{3} & \sqrt{2}\pi^+&\sqrt{2}K^+ \\
			\sqrt{2}\pi^- &-\pi^0+\eta/\sqrt{3}&\sqrt{2}K^0 \\
			\sqrt{2}K^-&\sqrt{2}\overline{K}^0&-2\eta/\sqrt{3}
		\end{pmatrix}\right) \, , 
	\end{equation}
 	and  \begin{flalign*}\label{ChargeMatrixShifted}
 		c&=\begin{pmatrix}
 			C_u-Q^a_u&0&0\\
 			0&C_d-Q^a_d&0\\
 			0&0&C_s-Q^a_s
 		\end{pmatrix}\\ \refstepcounter{equation}\tag{\theequation}\\
 		&=\frac{1}{3}\text{Tr}(c)\mathbf{1}+\frac{1}{2}\text{Tr}(c\lambda_3)\lambda_3+\frac{1}{2}\text{Tr}(c\lambda_8)\lambda_8=c_a\lambda_a \,, 
 	\end{flalign*}
 	is the shifted axial coupling matrix, which can be decomposed as a linear combination of Gell-Mann matrices and the unit matrix.
	After choosing $Q_a$ according to Eq. \eqref{MatrixQa}, the whole axion dependence is moved to the axial current and the mass terms. 	
	One can find the axion mass in 3-flavor ChPT in leading order in $w$ as
	\begin{equation}
		m_a^2=\frac{f_\pi^2m_\pi^2 (4z(1+z)+w(1-z)^2)}{4f_a^2 (1+z)^2(1+z+w)} \, .
	\end{equation}

	\subsection{Coupling to Pions}
	One can show that the trace argument of axial current for pions, keeping up to 3 fields, is given by \begin{equation}
		U\partial^\mu U^\dagger-U^\dagger \partial^\mu U=\frac{4 i}{3 f_\pi^3}\bigg( \partial_\mu\pi^b (\pi \pi)-\pi^b(\pi \partial_\mu\pi) \bigg)\lambda^b \, , 
	\end{equation}
	where now the index $b$ is restricted to $SU(3)$ generators contracted with pions, that is, $b=1,2,3$. Using $\text{Tr}\lambda^a\lambda^b=2\delta^{ab}$, axion-pion couplings are given by\begin{equation}
		\mathcal{L}_\text{axial}=-\frac{2}{3}\frac{\partial_\mu a}{f_af_\pi}c_a \bigg( \partial_\mu\pi^b (\pi \pi)-\pi^b(\pi \partial_\mu\pi) \bigg) \delta^{ab} \, .
	\end{equation}
	Taking into account the Kronecker-$\delta$ with restricted index $b$ and Eq.~\eqref{ChargeMatrixShifted}, we have $c_a\delta_{a3}=c_3$,  which can be further evaluated to $\text{Tr}(c\lambda_3)/2=(C_u-Q^a_u-C_d+Q^a_d)/2$.	
	Note that these terms are independent of $C_s$. After further simplifications we end up with \begin{equation} \label{eq::AxionPionLag}
		\mathcal{L}_{\text{axial}}=-\frac{C_u-C_d-Q^a_u+Q^a_d}{3f_\pi f_a}\big(\partial_\mu a\big)\bigg( 2\partial_\mu \pi_0 \pi_+ \pi_-  -   \pi_0 \partial_\mu \pi_+ \pi_-  -   \pi_0  \pi_+ \partial_\mu \pi_- \bigg) \, .
	\end{equation}
	Identifying the prefactor rescaled by $f_\pi f_a$ as the axion-pion coupling and using the explicit form of $Q_a$ from Eq. \eqref{MatrixQa}, we obtain the known result \cite{Chang:1993gm}  \begin{equation}
		C_{\pi}=-\frac{1}{3}\bigg(C_u-C_d-\frac{1-z}{1+z+w}\bigg) \, . 
	\end{equation}
	In astrophobic axion models $C_u=2/3$, $C_d=1/3$ and for approximate values $z \approx 1/2$ and $w \approx 0$ the pion coupling vanishes. Since scatterings with pions are the leading contribution to the abundance of thermal axions in generic models, $z\neq1/2$ and $w\neq 0$ may play a role in determination of $\dNeff$. However, as we have checked for the parameter space consistent with astrophysical constraints on the axion-nucleon, the production via scatterings with pions always contributes to $\dNeff$ less than $0.01$, far beyond the reach of future experiments.

	\subsection{Coupling to Kaons}\label{AppendixKaonCoupling}
	In the physical basis, with the choice of $Q_a$ given by Eq. \eqref{MatrixQa}, we move the axion interaction (up to order $1/f_a^2$ corrections) to the axial current. The only non-zero contribution to the axion kaon coupling comes from the axial current in Eq.~\eqref{ChPTLag} and for values $z=1/2$, $w=0$ has the form 
	\begin{equation}\label{eq::KaonPionAxionLagAllCouplings}\begin{split}
		\mathcal{L}_{aK}=-\frac{\partial_\mu a}{36 f_a f_\pi}&\bigg( -3\Big(C_u-C_d-\frac{1}{3}\Big)\Big(3\sqrt{2}K^+\pi^-\partial^\mu \bar{K}^0-3\sqrt{2}K^0\pi^+\partial^\mu K^- +\bar{K}^0\pi^0\partial^\mu K^0 \\ &-3\sqrt{2}\bar{K}^0 \pi^- \partial^\mu K^+ -2\bar{K}^0 K^0 \partial^\mu\pi^0+3\sqrt{2}K^- \pi^+ \partial^\mu K^0+K^-\pi^0\partial^\mu K^+\\& -2 K^- K^+ \partial^\mu \pi^0 +\pi^0 K^0 \partial^\mu\bar{K}^0+\pi^0 K^+\partial^\mu K^- \Big) \\& +3\Big(C_u+C_d-1-2C_s\Big)\Big(K^0\pi^0 \partial^\mu\bar{K}^0-\sqrt{2}K^+\pi^- \partial^\mu\bar{K}^0 -K^+\pi^0 \partial^\mu K^- \\&+\sqrt{2}K^0\pi^+  \partial^\mu K^- + \bar{K}^0\pi^0 \partial^\mu K^0 -\sqrt{2}\bar{K}^0 \pi^-  \partial^\mu K^+ -2 \bar{K}^0 K^0 \partial^\mu \pi^0\\&+2\sqrt{2}\bar{K}^0 K^+ \partial^\mu\pi^- -\sqrt{2} K^-\pi^+ \partial^\mu K^0 +K^-\pi^0 \partial^\mu K^+ -2K^-K^+\partial^\mu\pi^0 \\&-2\sqrt{2}K^- K^0 \partial^\mu\pi^+ \Big)\bigg) \, .
	\end{split}
	\end{equation}
	 For astrophobic axions with charges $C_u=2/3$, $C_d=1/3$, $C_s\neq 0$, for which the charge matrix Eq. \eqref{ChargeMatrixShifted} is $c=\text{diag}(0,0,C_s)$, only the contribution from strange quark coupling remains
	\begin{flalign*}\label{eq::KaonPionAxionLag}
		\mathcal{L}_{aK}=\frac{C_s}{6f_a f_\pi}&\partial_\mu a \bigg( \overline{K}^0(\pi^0 \partial^\mu K^0 -2K^0\partial^\mu\pi^0 -\sqrt{2}\pi^-\partial^\mu K^+ +2\sqrt{2}K^+\partial^\mu\pi^-)- \\ 
		&-K^- (\sqrt{2}\pi^+\partial^\mu K^0-2\sqrt{2}K^0 \partial^\mu\pi^++\pi^0\partial^\mu K^+-2K^+ \partial^\mu \pi^0)+\refstepcounter{equation}\tag{\theequation}\\&+K^0\pi^0 \partial^\mu \overline{K}^0-\sqrt{2}K^0 \pi^+ \partial^\mu K^--\sqrt{2}K^+\pi^- \partial^\mu \overline{K}^0-K^+\pi^0 \partial^\mu K^-  \bigg) \, .
	\end{flalign*}
	Note that those couplings are not suppressed in contrast  to the axion-pion coupling. As a result, the suppression of the kaon scattering rate comes solely from the Boltzmann factors and is insufficient to completely remove the kaon contribution to axion thermalization. The axion also couples to one $\eta$ and two kaons, but the scattering rate is  further suppressed by Boltzmann factors.
	
\section{Axion Production Rates}
\label{rates}

In this appendix we collect useful expressions for calculating thermal axion production rates, mainly following Ref.~\cite{DEramo:2017ecx}. 
\setcounter{equation}{0}
\setcounter{table}{0}

\subsection{Equilibrium Number Densities}
The number density of particles in thermal equilibrium is given by
\begin{equation}
	n_i^{\textrm{eq}}=g_i \int \frac{d^3 k}{(2\pi^3)} f^{\rm eq}_i \simeq \frac{g_i}{2\pi^2}m_i^2 T K_2\left( \frac{m_i}{T}\right),
\end{equation}
where $g_i$ denotes the number of internal degrees of freedom and  $f_i^{\rm eq} = 1/(e^{- E_i/T} \pm 1)$ denotes the phase space distributions with $E_i=  (\vec{k}^2+m_i^2)^{1/2}$.
One can approximate the equilibrium distributions for both bosons and fermions by a Maxwell-Boltzmann distribution, $f_i^\textrm{eq}\simeq \exp \left( -E_i/T\right)$. This allows to approximate the number densities by the last equality, where $K_2(m_i/T)$ denotes the modified Bessel function of second kind with  asymptotic behaviour
\begin{align}
K_2 (x) & = \begin{cases} 2/x^2 & x \ll 1 \\ e^{-x} \sqrt{\pi/ 2 x} & x \gg 1 \end{cases} \, .
\end{align}
The axion mass is negligible while being thermally produced. We use the exact Bose-Einstein distribution with equilibrium density
\begin{align}
n_a^{\rm eq} & = \frac{\zeta(3) }{\pi^2} T^3 \, .
\end{align}

\subsection{Collision Operators and Production Rates}

Axion production rates are related to the collision operator in the Boltzmann equation by $\Gamma_i (T) = {\cal C}_i/n^{\rm eq}_a$,  where $i$ collectively denotes  scattering processes $p_1p_2\leftrightarrow p_3 a$ and decays $p_1\leftrightarrow p_2 a$. For scatterings the collision operator given by (neglecting Pauli blocking)
\begin{equation}
		\mathcal{C}_{p_1p_2\leftrightarrow p_3 a}=\int d\Pi_1 d\Pi_2d\Pi_3d\Pi_a f_1^{\text{eq}}f_2^{\text{eq}} (2\pi)^4 \delta^4(p_1+p_2-p_3-p_a) |\mathcal{M}_{p_1p_2\rightarrow p_3 a}|^2 \, ,
	\end{equation}
	where $p_i$ denotes to the momentum of the $i$th particle, $d\Pi_i = d^3 p_i/ 2 E_i (2 \pi)^3$ is the Lorentz invariant phase space measure, and $|\mathcal{M}_{p_1p_2\rightarrow p_3 a}|^2$ denotes the squared matrix element of the scattering process including sums over initial and final polarizations (no averaging). The collision operator  can be related to the cross section 
	 \begin{equation}
		\sigma_{p_1p_2\rightarrow p_3 a}=\frac{1}{g_1 g_2}\frac{1}{4 p_1\cdot p_2 v_{12}}\int d\Pi_3d\Pi_a(2\pi)^4 \delta^4(p_1+p_2-p_3-p_a)  |\mathcal{M}_{p_1p_2\rightarrow p_3 a}|^2 \, ,
	\end{equation}
	where $g_1, g_2$ are the internal degrees of freedom of the initial particles and $v_{12}$ is their Lorentz invariant relative velocity, given by 
\begin{equation}
		v_{12}=\frac{\sqrt{(p_1\cdot p_2)^2-m_1^2m_2^2}}{2 p_1\cdot p_2}=\frac{s}{2 p_1\cdot p_2} \lambda^{1/2} \left(\frac{m_1}{\sqrt{s}},\frac{m_2}{\sqrt{s}}\right) \, ,
	\end{equation}
	with  $\lambda(x,y)=[1 -(x +y)^2][1 -(x-y)^2 ]$. This gives
	\begin{equation}
		\mathcal{C}_{p_1p_2\leftrightarrow p_3 a}= 2 g_1 g_2 \int d\Pi_1 d\Pi_2  f_1^{\text{eq}}f_2^{\text{eq}}  \lambda^{1/2}(x_1, x_2) s \sigma_{p_1p_2\rightarrow p_3 a} (s) \, , 
	\end{equation}
where $x_i = m_i/\sqrt{s}$ and the dependence on initial state momenta is encoded in the  center-of-mass energy $s$. Changing integration variables and performing one integration in the Boltzmannian limit $f_i^\textrm{eq}\simeq \exp \left( -E_i/T\right)$ leaves a single integral over the center-of-mass energy
\begin{equation}
		\mathcal{C}_{p_1p_2\leftrightarrow p_3 a}= \frac{ g_1 g_2}{32 \pi^4} T \int_{s_{\rm min}}^\infty \lambda(x_1,x_2) s^{3/2} \sigma_{p_1p_2\rightarrow p_3 a} (s) K_1 \left(\frac{\sqrt{s}}{T}\right) \, ds \, , 
		\label{Cscat}
	\end{equation}
 where $s_{\rm min} = {\rm max} \{ (m_1 + m_2)^2, (m_3 + m_a)^2 \}$, and $K_1 (x)$ denotes the modified Bessel function of second kind with  asymptotic behaviour
\begin{align}
K_1 (x) & = \begin{cases} 1/x & x \ll 1 \\ e^{-x} \sqrt{\pi/ 2 x} & x \gg 1 \end{cases} \, .
\end{align}
This collision term is conveniently rewritten in terms of the thermally averaged cross-section as
\begin{align}
\mathcal{C}_{p_1p_2\leftrightarrow p_3 a} = n_1^{\rm eq} n_2^{\rm eq} \langle \sigma_{p_1p_2\rightarrow p_3 a} v \rangle   \, , 
\end{align}
which defines the latter as (in the Boltzmannian limit)
	\begin{align} \label{eq::ThermAvgXSec} 
			\langle \sigma_{p_1p_2\rightarrow p_3 a} v \rangle= \frac{1}{8 m_1^2 m_2^2 T K_2(m_1/T)K_2(m_2/T)} \int_{s_{\rm min}}^\infty  \lambda(x_1,x_2) s^{3/2} \sigma_{p_1p_2\rightarrow p_3 a} (s) K_1 \left(\frac{\sqrt{s}}{T}\right) \, ds \, .
	\end{align}
If one of the initial particles (particle 2) can be considered massless to good approximation, $m_2 \to 0$, this simplifies to 
\begin{equation} \label{eq:ThermAvgXSecMassless}
		\langle \sigma_{p_1p_2\rightarrow p_3 a} v \rangle |_{m_2 = 0}=\frac{1}{16 m_1^2 T^3 K_2(m_1/T)}\int_{s_{\rm min}}^\infty  \left(1- \frac{m_1^2}{s} \right)^2 s^{3/2}  \sigma_{p_1p_2\rightarrow p_3 a} (s) K_1 \left(\frac{\sqrt{s}}{T}\right) \, ds \, .
	\end{equation}	
If the initial particles have the same mass, one has  
\begin{equation} \label{eq:ThermAvgXSecSameMass}
		\langle \sigma_{p_1p_2\rightarrow p_3 a} v \rangle  |_{m_2 = m_1} =\frac{1}{8 m_1^4  T K_2(m_1/T)^2 } \int_{s_{\rm min}}^\infty  \left( 1 - \frac{4 m_1^2}{s} \right) s^{3/2} \sigma_{p_1p_2\rightarrow p_3 a} (s) K_1 \left(\frac{\sqrt{s}}{T}\right) \, ds  \, .
	\end{equation}
	The axion production rate from scattering processes is finally given by 
	\begin{equation} \label{eq::RateScattering}
		\Gamma_{p_1p_2\rightarrow p_3 a}(T) =\frac{n_1^{\text{eq}}n_2^{\text{eq}}}{n_a^{\text{eq}}
		} \langle \sigma_{p_1p_2\rightarrow p_3 a} v \rangle \, ,
	\end{equation}
	and for the total production rate from scattering $\Gamma_S(T)$  all processes have to be summed 
	\begin{equation} \label{eq::RateScatteringTOT}
	\Gamma_S(T) = \sum_{p_1, p_2}	\Gamma_{p_1p_2\rightarrow p_3 a}(T)  \, .
	\end{equation}
	In case of axion production from decays, the same steps yield  for the collision term
	\begin{align}
			\mathcal{C}_{p_1\rightarrow p_2 a}&=\int d\Pi_1 d\Pi_2d\Pi_af_1^{\text{eq}} (2\pi)^4 \delta^4(p_1-p_2-p_a) |\mathcal{M}_{p_1\rightarrow p_2 a}|^2 \nonumber \\ &= g_1 \Gamma_{p_1\rightarrow p_2 a} \int \frac{d^3p}{(2\pi)^3}\frac{m_1}{E_1}f_1^{\text{eq}} \nonumber \\
			&= n^{\rm eq}_1  \Gamma_{p_1\rightarrow p_2 a} \frac{K_1 (m_1/T)}{K_2 (m_1/T)} \, , 
			\label{Cdecay}
	\end{align}
	where $|\mathcal{M}_{p_1 \to p_2 a}|^2$  is the squared matrix element of the decay process including sums over initial and final polarizations (no averaging), and $\Gamma_{p_1\rightarrow p_2 a}$ denotes the decay rate in the rest frame of particle 1, given by
	\begin{align}
	\Gamma_{p_1\rightarrow p_2 a} = \frac{1}{16 \pi g_1 m_1} |\mathcal{M}_{p_1 \to p_2 a}|^2 \lambda^{1/2} \left(\frac{m_2}{m_1}, \frac{m_a}{m_1}\right) \, . 
	\end{align}
If the mass of the axion is neglected, this simplifies to 
	 \begin{align}
	\Gamma_{p_1\rightarrow p_2 a} = \frac{1}{16 \pi g_1 m_1} |\mathcal{M}_{p_1 \to p_2 a}|^2 \left(1- \frac{m_2^2}{m_1^2} \right) \, . 
	\end{align}
The  axion production rate from decays is finally given by \begin{equation}\label{Eq:DecayRate}
		\Gamma_{p_1\rightarrow p_2 a} (T)= \frac{n_1^{\text{eq}}}{n_a^{\text{eq}}}\Gamma_{p_1\rightarrow p_2 a}\frac{K_1(m_1/T)}{K_2(m_1/T)} \, ,
	\end{equation}
	and for the total production rate from decays all processes have to be summed
	\begin{equation}\label{Eq:DecayRate}
		\Gamma_D (T)  = \sum_{p_1} \Gamma_{p_1\rightarrow p_2 a} (T) \, ,
	\end{equation}
	including charge multiplicities, e.g. for a single charged lepton $p_1 = \ell^+, \ell^-$. 
	\subsection{Pion and Kaon Scatterings}
	Scattering with pions is usually the main channel of axion production. Given the coupling Eq. \eqref{eq::AxionPionLag}, the known amplitude for three processes $\pi^0\pi^\pm \rightarrow a\pi^\pm$, $\pi^+\pi^- \rightarrow a\pi^0$ evaluates to \cite{Hannestad:2005df} \begin{equation}
		\sum_\text{processes} |\mathcal{M}|_{\pi\pi\rightarrow a\pi}^2=\frac{9}{4}\left( \frac{C_{\pi}}{f_\pi f_a}\right)^2 \left(s^2+t^2+u^2-3m_\pi^4\right).
	\end{equation}
	Neglecting the mass splitting between charged and neutral pions, we obtain a total thermally averaged cross section for all three processes using Eq. \eqref{eq::ThermAvgXSec} and Eq.\eqref{eq::RateScattering}. However, because ChPT is only EFT the integral in Eq. \eqref{eq::ThermAvgXSec} should only be performed up to cut-off $\sqrt{s}=\Lambda_\text{ChPT}=4\pi f_\pi$. Otherwise, artificial contributions from  divergent cross section would alter high-temperature behaviour of scattering rate. 
	
	Since pions are present in the thermal bath only after the QCDPT, we include this contribution in Boltzmann equation only for $T<T_\text{QCD}=158$~MeV \cite{Borsanyi:2020fev} by adding the Heaviside theta function $\theta(T_\text{QCD}-T)$. It is worth mentioning that working in LO of ChPT has been shown to be unreliable at temperature above $T\simeq70$~MeV \cite{DiLuzio:2021vjd,DiLuzio:2022gsc}. This contribution has been recently calculated precisely using phenomenological cross section \cite{Notari:2022ffe}. However, the LO ChPT suffices to show that in models considered in this work, pions give negligible contribution to the $\dNeff$.
	
	The scattering rate for processes involving kaons and pions $\pi K\rightarrow K a$ have been calculated in case of hadronic axions \cite{Notari:2022ffe}. In nucleophobic models these scattering rates are suppressed unless $C_s\neq0$, as explained above Eq. \eqref{eq::KaonPionAxionLag}. Given interactions Eq. \eqref{eq::KaonPionAxionLagAllCouplings} we used FeynRules \cite{Alloul:2013bka} and FeynCalc \cite{Shtabovenko:2016sxi,Shtabovenko:2020gxv} to derive the scattering amplitude for 8 processes: $\pi^0 K^\pm, K^0 \pi^\pm\rightarrow a \pi^\pm$; $\pi^0K^0, \pi^-K^+\rightarrow aK^0$; $\pi^+K^-, \pi^0 \overline{K}^0 \rightarrow a \overline{K}^0$, which for $w=0,z=1/2$ is given by a simple formula \begin{equation}
		\sum_\text{processes} |\mathcal{M}|_{\pi K\rightarrow  a K}^2=\left(\frac{\sqrt{3}C_s}{2f_\pi f_a}\right)^2t^2.
	\end{equation}

	We neglect the mass splitting of kaons, and pions using formulas Eq. \eqref{eq::ThermAvgXSec} and Eq. \eqref{eq::RateScattering} to obtain the rates. We take into account production from those scatterings below the QCDPT by adding the Heaviside theta function to the rate. The suppression of this rate with respect to the pion scatterings rate comes mostly from the Boltzmann factors. 
	
	Since LO of ChPT breaks down around $T\sim 70$ MeV, our results should be taken with a grain of salt. The calculation of the scatterings rate in NLO ChPT is beyond the scope of this paper, and we leave it for future work.
	
	\subsection{Lepton flavor-conserving Scatterings}
	
	There are two leptonic processes  that thermalize axions and conserve flavor - lepton annihilation ${\ell}^+{\ell}^-\rightarrow \gamma a$ and Compton-like scattering with axions at the final state ${\ell}^\pm \gamma\rightarrow {\ell}^\pm a$. 
	The resulting production rates are well-known, and for completeness we quote the results. 
	
	Two tree-level diagrams, $u$- and $t$-channel, contribute to process ${\ell}^+{\ell}^-\rightarrow \gamma a$. Upon phase space integration, the cross section as a function of center of mass energy is given by \cite{DEramo:2018vss}
	\begin{align}
		\sigma_{{\ell}^+{\ell}^-\rightarrow \gamma a}(s)=\frac{e^2 C_{\ell}^2 m_{\ell}^2}{4\pi f_a^2 (s-4m_{\ell}^2)} \tanh^{-1}\left( \sqrt{1-\frac{4m_{\ell}^2}{s}}\right) \, .
	\end{align}
	Two diagrams ($s$- and $u$-channel)  contribute to the cross section for Compton-like scattering  ${\ell}^\pm \gamma\rightarrow {\ell}^\pm a$, which reads~\cite{DEramo:2018vss}
	\begin{align} \label{eq:leptonphoton_xsec}
		\sigma_{{\ell}^\pm \gamma\rightarrow {\ell}^\pm a}(s)=\frac{C_{\ell}^2m_{\ell}^2e^2}{32 \pi f_a^2}\frac{2s^2\log(s/m_{\ell}^2)-3s^2+4m_{\ell}^2s-m_{\ell}^4}{s^2(s-m_{\ell}^2)} \, .
	\end{align}	
For large $s$ both scattering processes have the same scaling $\sigma \sim m_{\ell}^2/f_a^2 \log s/s$. This gives the scaling of the thermally averaged cross-sections in Eq.~\eqref{eq:ThermAvgXSecMassless} for large $T$ as 
\begin{align}
\langle \sigma v \rangle_{{\ell}^+{\ell}^-\rightarrow \gamma a}  & \xrightarrow{T \, \gg \, m_\ell}  \frac{C_\ell^2 m_\ell^2 e^2}{32 \pi f_a^2 T^2} \log \frac{2  T}{m_\ell} \, , \\
\langle \sigma v \rangle_{{\ell}^\pm \gamma\rightarrow {\ell}^\pm a}  & \xrightarrow{T \, \gg \, m_\ell}  \frac{C_\ell^2 m_\ell^2 e^2}{64 \pi f_a^2 T^2}  \log \frac{2  T}{m_\ell}  \, ,
\end{align}  
so that the production rates at high temperatures are
\begin{align}
\Gamma_{{\ell}^+{\ell}^-\rightarrow \gamma a} (T)  & \xrightarrow{T \, \gg \, m_\ell}  \frac{C_\ell^2 m_\ell^2 e^2 T }{8 \pi^3 \zeta(3) f_a^2}  \log \frac{2  T}{m_\ell}  \, , \\
\Gamma_{{\ell}^\pm \gamma\rightarrow {\ell}^\pm a}  (T)  & \xrightarrow{T \, \gg \, m_\ell}  \frac{C_\ell^2 m_\ell^2 e^2 T}{16 \pi^3 \zeta(3) f_a^2}  \log \frac{2  T}{m_\ell}  \, . 
\end{align}  
The total axion production rate from scatterings at high temperatures is then given by 
\begin{align}
\label{GShigh}
\Gamma_S (T)  & =  \Gamma_{{\ell}^+{\ell}^-\rightarrow \gamma a} (T)  +  \Gamma_{{\ell}^+ \gamma\rightarrow {\ell}^+ a}  (T) + \Gamma_{{\ell}^- \gamma\rightarrow {\ell}^- a}  (T) \nonumber \\
& \xrightarrow{T \, \gg \, m_\ell} \frac{C_\ell^2 \alpha  m_\ell^2 T}{ \pi^2 \zeta(3) f_a^2}  \log \frac{2  T}{m_\ell}  \, . 
\end{align}  
In the freeze-in regime axion production is dominated by temperatures slightly below $m_\ell$, and we approximate the scattering rate in this regime simply by taking the high-temperature expression above without the log but with a Boltzmann suppression factor, i.e.
\begin{align}
\label{GSm}
\Gamma_S (T) 
& \xrightarrow{T \, \lesssim \, m_\ell} \frac{C_\ell^2 \alpha m_\ell^2  T}{ \pi^2 \zeta(3) f_a^2} e^{-\frac{m_\ell}{T}} \, . 
\end{align}  
Instead for very low temperatures the scaling of the two processes with $T$ is different
\begin{align}
\langle \sigma v \rangle_{{\ell}^+{\ell}^-\rightarrow \gamma a}  & \xrightarrow{T \, \ll \, m_\ell}  \frac{C_\ell^2 e^2}{8 \pi f_a^2} \, , \\
\langle \sigma v \rangle_{{\ell}^\pm \gamma\rightarrow {\ell}^\pm a}  & \xrightarrow{T \, \ll \, m_\ell}  \frac{C_\ell^2 e^2 T^2}{ \pi m_\ell^2 f_a^2 }  \, ,
\end{align}  
and the production rates at low temperatures are Boltzmann suppressed
\begin{align}
\label{GSlow1}
\Gamma_{{\ell}^+{\ell}^-\rightarrow \gamma a} (T)  & \xrightarrow{T \, \ll \, m_\ell}  \frac{C_\ell^2 e^2 m_\ell^3 }{16 \pi^2 \zeta(3) f_a^2}  e^{- \frac{2  m_\ell}{T}}  \, , \\
\label{GSlow2}
\Gamma_{{\ell}^\pm \gamma\rightarrow {\ell}^\pm a}  (T)  & \xrightarrow{T \, \ll \, m_\ell}  \sqrt{\frac{2T}{\pi m_\ell}}  \frac{C_\ell^2 e^2 T^3}{\pi^2 \zeta(3) f_a^2}  e^{- \frac{m_\ell}{T}} \, . 
\end{align}

	\subsection{Lepton flavor-violating Decays }
	The decay rate for the process ${\ell} \rightarrow \ell^{\prime} a$ is given by~\cite{DEramo:2018vss} \begin{align}
		\Gamma_{{\ell}^\pm \rightarrow \ell^{\prime \pm }a}=C_{\ell \ell^\prime}^2 \frac{m_{\ell}^3}{64\pi f_a^2}\bigg( 1-\frac{m_{\ell^\prime}^2}{m_{\ell}^2} \bigg)^3\approx C_{\ell \ell^\prime}^2 \frac{m_{\ell}^3}{64\pi f_a^2} \, ,
	\end{align}
	where the last approximation holds to good approximation as  lepton masses are strongly hierarchical. The axion production rate from decays can be computed  using Eq. \eqref{Eq:DecayRate}, which gives\footnote{Flavor-violating scatterings $ {\ell} \gamma\rightarrow \ell^{\prime} a$ and $ {\ell} \ell^{\prime}\rightarrow  \gamma a$ have infrared divergence, which should cancel against real and virtual corrections to the tree-level decay rate. This has been demonstrated for the case of  $ {\ell} \gamma\rightarrow \ell^{\prime} a$ in Ref.~\cite{Czarnecki:2011mr}, and we expect a similar behaviour for $ {\ell} \ell^{\prime}\rightarrow  \gamma a$, at least for small $T \lesssim m_{\ell}$. Therefore these processes should only give sub-leading contribution to the axion production rate (peaked at $T \lesssim m_\ell/3$) and thus are omitted here.}   
\begin{align}
\Gamma_D (T)  = \frac{g_1}{2 \pi^2} m_\ell^2 T K_1 (m_\ell/T) \left( \Gamma_{{\ell}^+ \rightarrow \ell^{\prime + }a} + \Gamma_{{\ell}^- \rightarrow \ell^{\prime - }a} \right) \, ,
\end{align}  	
with $g_1 =2 $ for spin degrees of freedom. For large temperatures this becomes \begin{align}
\label{GDhigh}
\Gamma_D (T)   \xrightarrow{T \, \gg \, m_\ell}   \frac{C_{\ell \ell^\prime}^2 m_{\ell}^4}{32 \pi  \zeta(3) f_a^2 T}  \, ,
\end{align}	
When $T$ drops below the lepton mass $m_{\ell}$, the production rate becomes Boltzmann suppressed, giving
\begin{align}
\label{GDm}
\Gamma_D (T)   \xrightarrow{T \, \lesssim \, m_\ell}   \frac{C_{\ell \ell^\prime}^2 m_{\ell}^4}{32 \pi  \zeta(3) f_a^2 T} e^{-\frac{m_\ell}{T}}\, , 
\end{align}  
while for very low temperatures one obtains 
\begin{align}
\label{GDlow}
\Gamma_D (T)   \xrightarrow{T \, \ll \, m_\ell}   \sqrt{\frac{2T}{ \pi m_\ell}} \frac{C_{\ell \ell^\prime}^2 m_{\ell}^5}{64   \zeta(3) f_a^2 T^2}  e^{-\frac{m_\ell}{T}}\, .
\end{align}

\section{Boltzmann Equation}
\label{Boltzmann}

\setcounter{equation}{0}
\setcounter{table}{0}	
	
The number density $n_a$ of axions is governed by the integrated Boltzmann equation \cite{Cadamuro:2010cz} \begin{align}
		\frac{dn_a}{dt}+3Hn_a=\bigg( \sum_i \Gamma_i\bigg) \bigg( n_a^\eq-n_a \bigg) \, ,
	\end{align}
	where $n_a^{\text{eq}}$ is the number density of axions at equilibrium, $H$ is the Hubble parameter 
\begin{align}
H = \frac{T^2}{M_{\rm Pl}} 1.66 \sqrt{g_* (T)}  \, , 
\end{align}	
with 	$g_* (T)$ denoting the total number of relativistic  degrees of freedom
and $\Gamma_i$ are the single axion production rates considered above. It is convenient to work with dimensionless temperature variables $x=m/T$ and the yields $Y_a=n_a/s$, where $m$ is chosen as the mass of the heaviest particle involved in the production process  (in our calculations we use $m=m_\tau$) 
	and $s$ is the entropy density \begin{align}
		s=\frac{2\pi^2  T^3}{45}g_{*s}(T) \, ,
	\end{align}
	where $g_{*s} (T)$ is the effective number of relativistic entropy degrees of freedom\footnote{We use the results from Ref.~\cite{Saikawa:2018rcs} for both $g_{*s} (T)$ and $g_{*}(T)$,  but our results are not very sensitive on the exact values of $g_{*}$ we use, which has already been noted in Ref.~\cite{DEramo:2018vss}.} It is clear that $Y_a^\eq$ remains constant as long as $g_{*s}$ does not change. Since axions decouple around the QCDPT, there is a rapid change in $g_{*s}$, and accordingly the equilibrium yield changes as well. 
	
	Using entropy conservation, $d (sa^3)/dt = 0$, we can express the time-derivative as
	\begin{align}
		\frac{dx}{dt}=xH \bigg( \frac{T}{3s}\frac{ds}{dT}\bigg)^{-1} =xH \bigg(1-\frac{x}{3g_{*s}}\frac{dg_{*s}}{dx}\bigg)^{-1}\, .
	\end{align}
	The Boltzmann equation in new variables reads \begin{align}
		sHx\frac{dY_a}{dx}=\bigg( 1-\frac{x}{3g_{*s}}\frac{dg_{*s}}{dx}\bigg) n_a^\eq \sum_i \Gamma_i \bigg(1-\frac{Y_a}{Y_a^\eq}\bigg) \, .
		\label{eq::BEfinal}
	\end{align}
	We solve this equation numerically on the interval $x \in [0.01,190]$ assuming vanishing initial yield,  $Y_a^i=Y_a (x=0.01) =0$. Note that axion production from electron scattering is active at $T\sim m_e$, in this case we solve the Boltzmann equation up to $x=4000$. 
	 In order to improve the computation time, we evaluate the rates $\Gamma_i$ for logarithmically distributed points $x_i$ and use the spline interpolation to recover continuous functions.

	Thermal axions contribute to the total energy density of radiation, which is parametrized by the additional effective number of neutrinos $\Delta N_{\text{eff}}$ as
	\begin{align}
		\Delta N_{\text{eff}}=\frac{8}{7} \left( \frac{11}{4}\right)^\frac{4}{3} \frac{\rho_a}{\rho_\gamma}\bigg|_{T_\text{CMB}} \, , 
		\label{dNeffDefinition}
	\end{align}
	where $\rho_\gamma$ and $\rho_a$ are the energy densities of photons and axions, respectively. One can estimate $\rho_a$ in terms of the axion number density $n_a$, and $\rho_\gamma$ in terms of the entropy density $s$, obtaining~\cite{DEramo:2021lgb} \begin{align}
	\label{eq:dNeff_Ya}
		\dNeff= \frac{4}{7} \left( \frac{11}{4} \right)^\frac{4}{3} \left(\frac{2\pi^4}{45 \, \zeta(3) }g_{*s}(T_{\text{CMB}}) Y_a(T_{\text{CMB}}) \right)^\frac{4}{3} \, .
	\end{align}
	Although $\dNeff$ is set by the axion yield at $T_\text{CMB}$, we may  identify it with the asymptotic yield $Y_a\big|_{T_\text{CMB}}=Y_a^\infty$, which we numerically evaluate at the endpoint of the Boltzmann integration interval since at late times all axion interactions are frozen. If the yield of axions is small we can simply take $g_{*s}(T_{\text{CMB}})\simeq g_{*s}^\text{SM}(T_{\text{CMB}})=43/11$ and the numerical formula reads \begin{align}
		\Delta N_{\text{eff}}=74.85 \big(Y_a^\infty\big)^{\frac{4}{3}} \, . \label{eq:dNeffApprox}
	\end{align}
	However, if the abundance of axions is large, we cannot neglect their contribution to the number of entropic degrees of freedom and a better estimate reads~\cite{DEramo:2021lgb}
	\begin{align} 
		\dNeff=\frac{4}{7} \left( \frac{11}{4} \right)^\frac{4}{3} \left(\frac{\frac{2\pi^4}{45\zeta(3) }g_{*s}^{\text{SM}}(T_{\text{CMB}}) Y_a^\infty}{1-\frac{2\pi^4}{45\zeta(3) } Y_a^\infty} \right)^\frac{4}{3} \, .
		\label{eq:dNeffFinal}
	\end{align}
	We use this expression throughout our analysis, although its effect is non-negligible only for $\dNeff\gtrsim0.2$, and even then is merely a few-percent correction compared to Eq. \eqref{eq:dNeffApprox}.

We note that the estimate in Eq.~\eqref{eq:dNeffFinal} gives a good approximation only when the axion follows a thermal distribution, which is the case for freeze-out production. Instead for freeze-in production Eq.~\eqref{eq:dNeffFinal} underestimates $\dNeff$. This is because Eq.~\eqref{eq:dNeff_Ya} is based on the assumption that  $\rho_a = \pi^2 (\pi^2n_a/\zeta(3))^{4/3}/30$, for which the typical energy of axions is $\rho_a/n_a \sim n_{a}^{1/3}$. However, for freeze-in production $n_a^{1/3}$ is smaller than the actual typical energy $(n_{a}^{\rm eq})^{1/3} \sim T$ of thermally produced axions, so that the actual energy density of the axions is larger than obtained from Eq.~\eqref{eq:dNeff_Ya}.

Let us estimate how much we underestimate $\dNeff$. A more precise estimate of $\rho_a$ is
\begin{align}
\label{eq:rhoa_app}
 \rho_{a,1} = \rho_a^{\rm eq} \frac{n_a}{n_a^{\rm eq}}
\end{align}
at the temperature $T_{\rm FI}$ where the freeze-in production is peaked at. This relation is justified since typically $\dot{\rho}_a/\rho_a^{\rm eq} \simeq \dot{n}_a/n_a^{\rm eq}$. Taking the ratio with the energy density $\rho_{a,2}$ estimated by $\rho_{a,2} = \pi^2(\pi^2n_a/\zeta(3))^{4/3}/30$, we obtain
\begin{align}
\frac{\rho_{a,2}}{\rho_{a,1}} \simeq 0.5 \left(\frac{\dNeff}{0.05}\right)^{1/4} \left(\frac{g_{*s}(T_{\rm FI})}{10}\right)^{1/3},
\end{align}
where $\dNeff$ is obtained by Eq.~\eqref{eq:dNeffApprox}. One can see that the underestimation is more significant when $\dNeff$ is smaller and the freeze-in production is peaked at lower temperatures. The most underestimated channel is the scattering off muons, where $g_{*s}(T_{\rm FI})\simeq 10$. For the phenomenologically interesting range  $\dNeff > 0.05$, the underestimation is at  most off by a factor 2. This corresponds only to a 30\% change in $f_a$ since $\dNeff \propto f_a^{-8/3}$. For freeze-in production by other particles, where $g_{*s}(T_{\rm FI})$ is larger, the impact is even smaller.

Given that $\dNeff$ is appreciably underestimated only for the production off muons, we use Eq.~\eqref{eq:dNeffFinal} throughout the paper. Note that Eq.~\eqref{eq:rhoa_app} is also still approximate and requires the determination of $T_{\rm FI}$. A more exact computation will require the derivation of $\dot{\rho}_a$, which we leave for future work.

\subsection{Approximate Solution to Boltzmann Equation}
	It is possible to obtain an approximate analytic solution of the Boltzmann equation if one assumes approximately constant $g_{*s}$~\cite{Ferreira:2018vjj, DEramo:2018vss}, which allows to simplify $g_{*s}'(x)=0$ and take $Y_a^\eq$ constant.  Re-introducing the collision operators  ${\cal C}_i=n_a^\textrm{eq}\Gamma_i$  with $i$ denoting collectively scattering and decays, the Boltzmann equation~\eqref{eq::BEfinal} simplifies to 
\begin{align}
		\frac{dY_a}{1-\frac{Y_a}{Y_a^\eq}} = \frac{ \sum_i {\cal C}_i (x) x^4 dx}{H(m) s(m)} \, ,
		\label{eq:BEapp}
	\end{align}
 where we used the temperature scaling  of entropy density and Hubble rate scale  $H(x)=H(m)x^{-2}$, $s(x)=s(m)x^{-3}$ for constant $g_{*s}$.
With the initial condition $Y(x=0)=0$ one can integrate both sides to obtains the late-time yield $Y_a^{\infty} = Y_a (x \to \infty)$ as
 \begin{equation}
	Y_a^\infty =Y_a^\eq (m) \bigg(1-\exp\bigg[-\frac{\sum_i \int_0^\infty x^4 {\cal C}_i(x) dx} {H(m) s(m) Y^{\rm eq}_a (m) }\bigg]\bigg) \, .
	\label{BMsol}
\end{equation}
where we have evaluated for definiteness $Y_a^{\eq}$ at $T=m$. We now perform the remaining integral using the exact expressions in Eq.~\eqref{Cscat} and \eqref{Cdecay}. The latter gives (adding a factor 2 for charge multiplicities)
\begin{align}
{\cal C}_{p_1 \to p_2 a} (x) =   C_{\ell \ell^\prime}^2 \frac{m_\ell^6}{32 \pi^3 f_a^2}  \frac{K_1 (x)}{x} \, , 
\end{align}
and since 
\begin{align}
\int_0^\infty x^3 K_1(x) dx =  \frac{3 \pi}{2} \, , 
\end{align}
we obtain for decays 
\begin{align}
\int_0^\infty x^4 {\cal C}_{p_1 \to p_2 a} (x) =   C_{\ell \ell^\prime}^2 \frac{3 m_\ell^6}{64 \pi^2 f_a^2}  \,. 
\end{align}
Instead for scattering the collision operators are (including a factor of 2 for charge multiplicity in $ {\ell}^\pm \gamma\rightarrow {\ell}^\pm a$)
\begin{align}
		\mathcal{C}_{{\ell}^+{\ell}^-\rightarrow \gamma a} (x) & = \frac{m_\ell}{8 \pi^4 x}  \int_{4 m_\ell^2}^\infty \left(1-\frac{4 m_\ell^2}{s} \right) s^{3/2} \sigma_{{\ell}^+{\ell}^-\rightarrow \gamma a} (s) K_1 \left(\frac{x \sqrt{s}}{m_\ell}\right) \, ds \, , \nonumber 	\\
		& = \frac{4 m_\ell^6}{ \pi^4 x}  \int_0^\infty y \sqrt{1+y} \, \sigma_{{\ell}^+{\ell}^-\rightarrow \gamma a} (y) K_1 \left(2 x \sqrt{1+y} \right) \, dy \, ,
\end{align}	
where we substituted $s = 4 m_\ell^2 (1+y)$, and 	
\begin{align}
			\mathcal{C}_{{\ell}^\pm \gamma\rightarrow {\ell}^\pm a} (x) & = \frac{m_\ell}{4 \pi^4 x}  \int_{ m_\ell^2}^\infty \left(1-\frac{ m_\ell^2}{s} \right)^2 s^{3/2} \sigma_{{\ell}^\pm \gamma\rightarrow {\ell}^\pm a} (s) K_1 \left(\frac{x \sqrt{s}}{m_\ell}\right) \, ds \, , 	\nonumber \\
			& = \frac{ m_\ell^6}{4 \pi^4 x}  \int_0^\infty \frac{y^2}{ \sqrt{1+y}}  \sigma_{{\ell}^\pm \gamma\rightarrow {\ell}^\pm a} (y) K_1 \left( x \sqrt{1+y} \right) \, dy \, ,
\end{align}
where we substituted $s = m_\ell^2 (1+y)$. While these integrals are difficult, we can perform first the temperature (i.e. $x$) integration of the integrand, which leaves an expression that can be easily integrated over $y$ analytically, giving in total for scattering
\begin{align}
\int_0^\infty x^4 {\cal C}_{p_1 p_2 \to p_3 a} (x) =   C_{\ell \ell}^2 \frac{ \alpha m_\ell^6}{ \pi^2 f_a^2} \begin{cases} \frac{3}{32} & {\ell}^+{\ell}^-\rightarrow \gamma a \\ \frac{4}{21 \pi} & {\ell}^\pm \gamma\rightarrow {\ell}^\pm a \end{cases} \,. 
\end{align}

\subsection{Thermal Freeze-Out}

If interactions bring axions into thermal equilibrium at early times, i.e. $\Gamma_i (m)/H(m)\gg 1$, which can always be achieved for sufficiently   small $f_a$, the late-time abundance is dominated by the first term in Eq.~\eqref{BMsol}. This gives the yield as $Y_a^\infty=Y_a^\eq(T_d)$, where $T_d$ is the decoupling temperature defined by $H(T_d) = \Gamma_i(T_d)$, which enters the yield only through $g_{*s}(T_d)$. The latter decreases with $f_a$, so for sufficiently small $f_a$ the late-time yield reaches a plateau, and so does  $\dNeff$ according to Eq.~\eqref{eq:dNeffApprox}, or better \eqref{eq:dNeffFinal}.

We now estimate the decoupling temperature and the corresponding yield using the asymptotic forms of the scattering and decay rates. Since decoupling happens at moderately late times $x \approx m/10$, we use the asymptotic forms in the low-temperature regime Eq.~\eqref{GSm} and Eq.~\eqref{GDm}. The decoupling temperatures for scattering and decays are thus given by solving
\begin{align}
T^d_S & \approx m \log \left[ \frac{C^2 \alpha m^2 M_{\rm Pl} }{1.66 \sqrt{g_{*s} (T^d_S)} \pi^2 \zeta(3) f_a^2 T^d_S } \right] \, , \\
T^d_D & \approx m \log \left[ \frac{C^2 m^4 M_{\rm Pl} }{1.66 \sqrt{g_{*s} (T^d_D)} 32 \pi \zeta(3) f_a^2 (T^d_D)^3 } \right] \, .
\end{align}

\subsection{Thermal Freeze-In}
\label{FIapp}
When the interactions are too weak to bring axions to equilibrium, $\Gamma_i (m)/H(m) \ll 1$, which happens for sufficiently large $f_a$, one can expand the exponential n Eq.~\eqref{BMsol} to leading order obtaining 
\begin{align}
	Y_a^\infty &  \simeq Y_a^\eq (m) \frac{\sum_i \int_0^\infty x^4 {\cal C}_i(x) dx} {H(m) s(m) Y^{\rm eq}_a (m) } 
	 = \frac{m_\ell^6}{\pi^2 f_a^2 H(m) s(m)} \begin{cases} C_{\ell \ell}^2 \frac{3 \alpha}{32} & {\ell}^+{\ell}^-\rightarrow \gamma a  \\ C_{\ell \ell}^2 \frac{4 \alpha}{21 \pi}  &  {\ell}^\pm \gamma\rightarrow {\ell}^\pm a \\ 
C_{\ell \ell^\prime}^2 \frac{3}{64} 	& {\ell}^\pm \rightarrow {\ell}^{\prime \pm} a \end{cases} \nonumber \\
& = 1.4 \times 10^{-3} \frac{ \, m_\ell M_{\rm Pl}}{ f_a^2 g_{*s} (m_\ell) \sqrt{g_*(m_\ell)} } \begin{cases}  9.4  \alpha \, C_{\ell \ell}^2 & {\ell}^+{\ell}^-\rightarrow \gamma a  \\ 6.1 \alpha  \, C_{\ell \ell}^2 &  {\ell}^\pm \gamma\rightarrow {\ell}^\pm a \\ 
4.7 \, C_{\ell \ell^\prime}^2 	& {\ell}^\pm \rightarrow {\ell}^{\prime \pm} a \end{cases}  \, ,
\end{align}
so apart from the $\alpha$-suppression scattering and decays give similar contributions (see also Ref.~\cite{DEramo:2020gpr, Panci:2022wlc}). The yields scale as $m_\ell/f_a^2$, so that one obtains with Eq.~\eqref{eq:dNeffApprox} the scaling $\dNeff\propto f_a^{-8/3}$. These analytic formulas agree quite well with the numerical results (less so for muon scattering and decays, since the number of SM relativistic degrees of freedom changes rapidly around the muon mass).
\subsection{Transition Region}
We now estimate the temperature $T^{\rm eq}$ at which scatterings or decays bring axions into thermal equilibrium in the very early universe at $T \gg m$, which happens only for sufficiently small $f_a$. For this we evaluate the defining equations $H(T^{\rm eq}_i) =  \Gamma_i(T^{\rm eq}_i)$ in the high-temperature regime for the rates, using Eq.~\eqref{GShigh} and \eqref{GDhigh}. This leads to
\begin{align}
T^{\rm eq}_S & \approx  \frac{C_{\ell \ell}^2 \alpha m_\ell^2 M_{\rm Pl}}{ \pi^2 1.66  \zeta(3)  \sqrt{g_{* s}^{\rm UV}} f_a^2} \log \frac{2C_{\ell \ell}^2 \alpha m_\ell M_{\rm Pl}}{ \pi^2 1.66  \zeta(3)  \sqrt{g_{* s}^{\rm UV}} f_a^2} \, , \\
T^{\rm eq}_D & \approx  \left( \frac{C_{\ell \ell^\prime}^2 m_\ell^4 M_{\rm Pl}}{32 \pi 1.66  \zeta(3)  \sqrt{g_{* s}^{\rm UV}} f_a^2} \right)^{1/3} \, ,
\end{align}
where we worked at logarithmic accuracy in $T^{\rm eq}_S$ and approximated in both cases $g_{* s} (T^{\rm eq}_i) \approx g_{* s}^{\rm UV} \approx 104$, since $T^{\rm eq}_i \gg \TeV$. The point $T_i^{\rm eq} \approx m_\ell$ marks the transition regime between freeze-out and freeze-in, which we can estimate extrapolating the high-temperature expressions above. It  corresponds to the axion decay constant $(f_a/C_i)^{\rm eq}_i$ given by
\begin{align}
(f_a/C_{\ell \ell})^{\rm eq}_S & \approx \left( \frac{\alpha m_\ell M_{\rm Pl}}{ \pi^2 1.66  \zeta(3)  \sqrt{g_{* s}^{\rm UV}} } \log 2 \right)^{1/2} = 2 \times 10^7 \GeV \sqrt{\frac{m_\ell}{\GeV}} \, , \nonumber \\
(f_a/C_{\ell \ell^\prime})^{\rm eq}_D & \approx  \left( \frac{m_\ell M_{\rm Pl}}{32 \pi 1.66  \zeta(3)  \sqrt{g_{* s}^{\rm UV}} }  \right)^{1/2} = 8 \times 10^7 \GeV \sqrt{\frac{m_\ell}{\GeV}}\, .
\end{align}

\section{Astrophobic DFSZ Models}
\label{models}

\setcounter{equation}{0}
\setcounter{table}{0}

In this appendix we describe in detail the construction of astrophobic DFSZ models as SM extensions with two or three Higgs doublets. We begin with the general description of the quark Yukawa  sector and the resulting axion couplings, which are  determined by the PQ charges of Higgs doublets and flavor rotations. The form of the scalar  potential then fixes all charges in terms of the discrete choices for the Yukawa sector, the vacuum angles and flavor rotations. By scanning over these possibilities, we systematically identify all nucleophobic models, for models with two and three Higgs doublets. We finally include the charged lepton Yukawa sector and construct models with suppressed couplings to electrons, muons and/or photons.  

 \subsection{Quark Yukawa Sector}

To the SM fermion fields we add $n$ scalar Higgs doublets $h_i$ with hypercharge $Y=-1/2$ and a complex singlet scalar $\phi$. The quark Lagrangian is taken to be invariant under a $U(1)_{\rm PQ}$ symmetry, with the most general charge assignment consistent with a $2+1$ flavor structure, as shown in Table \ref{PQ}. 
\begin{table}[t]
\centering
\begin{tabular}{|c||cc|cc|cc|cc|cc||cc|}
\hline
 &  $q_{L3}$ & $q_{La}$ & $u_{R3}$ & $u_{Ra}$ & $d_{R3}$ & $d_{Ra}$ & $\ell_{L3}$ & $\ell_{La}$ & $e_{R3}$ & $e_{Ra}$ & $h_i$ &  $\phi$ \\
 \hline
 $U(1)_{\rm PQ}$ & $X_{q_3}$ & $X_{q_a}$ & $X_{u_3}$ & $X_{u_a}$ & $X_{d_3}$ & $X_{d_a}$ &  $X_{\ell_3}$ & $X_{\ell_a}$ & $X_{e_3}$ & $X_{e_a}$  & $X_i$  & $X_{\phi}$  \\
 \hline
 \end{tabular}
 \caption{\label{PQ} PQ charge assignments, $a = 1,2$ denotes the first two fermion generations, while $i = 0, \hdots, n-1$ runs over Higgs doublets.
 }
\end{table}
\noindent The general quark Yukawa Lagrangian reads 
\begin{align}
{\cal L} & = - y^{u}_{33} \overline{q}_{L3} u_{R3} h^u_{33} - y^{u}_{3a}  \overline{q}_{L3} u_{Ra} h^u_{3a} - y^{u}_{a3} \overline{q}_{La} u_{R3} h^u_{a3} - y^{u}_{ab}  \overline{q}_{La} u_{Rb} h^u_{ab} \nonumber \\
&  + y^{d}_{33} \overline{q}_{L3} d_{R3} \tilde{h}^d_{33} + y^{d}_{3a}  \overline{q}_{L3} d_{Ra} \tilde{h}^d_{3a} + y^{d}_{a3} \overline{q}_{La} d_{R3} \tilde{h}^d_{a3} + y^{d}_{ab} \overline{q}_{La} d_{Rb} \tilde{h}^d_{ab} + {\rm h.c.},   
\label{L1}
\end{align}
where $\tilde{h}_i = i \sigma^2 h_i^*$, $a,b = 1,2$ and each Higgs is chosen from the set of $n$ Higgs fields $h_i$, $i = 0 \hdots n-1$, e.g., $h^u_{3a} = h_2, h^d_{33} = h_0$, etc. Schematically, one has
\begin{align}
y_u & \sim \begin{pmatrix} h^u_{ab} & h^u_{a3} \\ h^u_{3a} & h^u_{33} \end{pmatrix} \, , &
y_d & \sim \begin{pmatrix} h^d_{ab} & h^d_{a3} \\ h^d_{3a} & h^d_{33} \end{pmatrix} \, , 
\end{align}
where we  indicate in {\bf 2} +{\bf 1} flavor space  to which Higgs field the respective quark bilinears couple to. Note that we require that all Yukawa couplings are allowed by the PQ symmetry, i.e., there are no Yukawa textures (for models where this assumption is relaxed see e.g., Ref.~\cite{Bjorkeroth:2018ipq}). This gives five constraints, which determines fermion charges in terms of Higgs charges, up to a single quark charge $X_{q_3}$%
\footnote{This follows from conserved  baryon number, which could be used to redefine $U(1)_{\rm PQ}$ such that  $X_{q_3} = 0$.} 
\begin{gather}
X_{u_a}  = - X_{h^u_{3a}}  + X_{q_3} \, ,   \qquad X_{d_a}  =  X_{h^d_{3a}}  + X_{q_3} \, ,  \qquad  X_{q_a}  =  - X_{h^u_{33}} + X_{h^u_{a3}} + X_{q_3}  \, \\ 
X_{u_3}  = - X_{h^u_{33}} + X_{q_3}\, ,   \qquad  X_{d_3}  =  X_{h^d_{33}} + X_{q_3} \,  .
\end{gather}
Moreover, there are three consistency conditions relating Higgs charges as
\begin{align}
 X_{h^u_{33}}   - X_{h^u_{a3}}  = X_{h^u_{3a}} - X_{h^u_{ab}} = X_{h^d_{a3}} - X_{h^d_{33}}   = X_{h^d_{ab}} - X_{h^d_{3a}}  \, .
\end{align}

\subsection{Axion Couplings}

The scalar potential is constructed with a single global $U(1)_{\rm PQ}$ symmetry, and suitable to generate vacuum expectation values for $\phi$ and all Higgs doublets $h_i$. The vacuum configuration breaks the global PQ symmetry spontaneously, and the corresponding Goldstone boson is the axion, which enters the Lagrangian as the phase of $\phi$ and Higgs fields.  Since Yukawas are PQ invariant, the axion couplings can be removed by performing the following flavor-diagonal fermion field redefinitions (which is just a local PQ transformation acting only on fermions) 
 \begin{align}
 f \to f \, e^{i X_f a (x) /v_{\rm PQ}}   \, .
 \end{align}
Since this transformation is anomalous, it generates axion couplings to gauge field strengths, and since it is local it modifies the fermion kinetic terms.  Here  $v_{\rm PQ}$ denotes the PQ breaking scale, which for $v_\phi \gg v_i$ is set by the singlet vev, $v_{\rm PQ} \approx X_\phi v_\phi$.

The  axion couplings to gluons and photons are given by
\begin{align}
{\cal L}_{\rm anom} = N \frac{a}{v_{\rm PQ}} \frac{\alpha_s}{4 \pi} G_{\mu \nu} \tilde{G}^{\mu \nu} + E \frac{a}{v_{\rm PQ}} \frac{\alpha_{\rm em}}{4 \pi} F_{\mu \nu} \tilde{F}^{\mu \nu} \, , 
\end{align}
with the dual field strength $\tilde{F}_{\mu \nu} = \frac{1}{2} \xi_{\mu \nu \rho \sigma} F^{\rho \sigma}, \xi^{0123} = -1$ and the anomaly coefficients
\begin{align}
\label{N}
2N & =  4 X_{q_a} -  X_{u_3} - 2 X_{u_a} - X_{d_3} - 2 X_{d_a} \nonumber \\
& =  2 X_{h^u_{ab}} +  X_{h^u_{33}} - 2 X_{h^d_{ab}}  - X_{h^d_{33}}   \, , \\
\nonumber \\
  E & = E_Q + E_L \nonumber \\
  & = \frac{5}{3} \left(   2 X_{q_a} + X_{q_3} \right) - \frac{4}{3}  \left( 2 X_{u_a} + X_{u_3}  \right) -  \frac{1}{3} \left(2 X_d+  X_{d_3}  \right) +E_L   \nonumber \\
  & =  \frac{8}{3} X_{h^u_{ab}} + \frac{4}{3} X_{h^u_{33}} - \frac{2}{3} X_{h^d_{ab}} - \frac{1}{3} X_{h^d_{33}}  + E_L   \, ,
 \end{align}
 where we included also a generic contribution from the charged lepton sector $E_L$, to be discussed below. 
 
From the kinetic terms one obtains axion-fermion couplings in the flavor interaction basis
\begin{align}
{\cal L} & =  \frac{\partial_\mu a}{v_{\rm PQ}} \left[ \overline{u}_{i} \gamma^\mu \left( \tilde{C}^q_{ij} P_L + \tilde{C}^u_{ij} P_R \right)  u_{j} + \overline{d}_{i} \gamma^\mu \left( \tilde{C}^q_{ij} P_L + \tilde{C}^d_{ij} P_R \right)  d_{j}  \right] \, , 
\end{align}
with
\begin{align}
\tilde{C}^q_{ij} & = \left(  X_{h^u_{33}} - X_{h^u_{a3}} - X_{q_3} \right) \delta_{ij} + {\rm diag} (0, 0,   X_{h^u_{a3}} -  X_{h^u_{33}} )  \, , \nonumber \\
\tilde{C}^u_{ij} & =   \left(  X_{h^u_{3a}}- X_{q_3}  \right) \delta_{ij} + {\rm diag} (0, 0,  X_{h^u_{33}} - X_{h^u_{3a}})  \, , \nonumber \\
  \tilde{C}^d_{ij} & = \left( - X_{h^d_{3a}}  - X_{q_3} \right) \delta_{ij} +  {\rm diag} (0, 0, X_{h^d_{3a}} - X_{h^d_{33}})  \,  .  
\end{align}
In the mass basis we finally obtain
\begin{align}
{\cal L} & = \frac{\partial_\mu a}{v_{\rm PQ}} \overline{u}  \gamma^\mu \left( C^{u_L} P_L + C^{u_R} P_R \right)  u +  \frac{\partial_\mu a}{v_{\rm PQ}}  \overline{d}  \gamma^\mu \left( C^{d_L}   P_L + C^{d_R}  P_R \right)  d \, , 
\end{align}
with
\begin{align}
C^{u_L}_{ij} & = \left(  X_{h^u_{33}} - X_{h^u_{a3}} - X_{q_3}  \right) \delta_{ij} -   \left(  X_{h^u_{33}} - X_{h^u_{a3}}  \right) \xi^{u_{L}}_{ij}    \, , \nonumber \\
C^{d_L}_{ij} & = \left(  X_{h^u_{33}} - X_{h^u_{a3}} - X_{q_3}  \right)  \delta_{ij} - \left(  X_{h^u_{33}} - X_{h^u_{a3}}  \right)  \xi^{d_{L}}_{ij}   \, , \nonumber \\
C^{u_R}_{ij} & =  \left(  X_{h^u_{3a}}- X_{q_3}  \right) \delta_{ij} + \left( X_{h^u_{33}} - X_{h^u_{3a}} \right) \xi^{u_{R}}_{ij} \, , \nonumber \\
C^{d_R}_{ij} & =  \left( - X_{h^d_{3a}}  - X_{q_3} \right)  \delta_{ij} + \left(    X_{h^d_{3a}} - X_{h^d_{33}} \right)\xi^{d_{R}}_{ij} \,   ,
 \label{Cij}
\end{align}
where the flavor structure is controlled by the matrices 
\begin{align}
\xi^{f_{P}}_{ij} \equiv (V_{fP})^*_{3i}  (V_{fP})_{3j} \, , \qquad f = u,d \, ,  \qquad P= L,R \, , 
\end{align}
 which depend on the unitary rotations $(V_{fP})_{ij}$ defined by $(V_{UL})^\dagger  m_u V_{UR} = m_u^{\rm diag}$ etc. They  satisfy
 \begin{align}
 0 & \le \xi^{f_P}_{ii} \le 1 \, , & \sum_i \xi^{f_P}_{ii}  & = 1 \, , &  |\xi^{f_P}_{ij}| & = \sqrt{\xi^{f_P}_{ii} \xi^{f_P}_{jj}} \, ,
 \end{align}
 and therefore depend only on 2 independent real parameters in each sector $u_L, u_R, d_R$.

Finally we adopt the standard convention for the axion decay constant $f_a =  v_{\rm PQ}/(2N)$, and write the Lagrangian as
\begin{align}
{\cal L} & = \frac{1}{2} (\partial_\mu a)^2 +  \frac{a}{f_a} \frac{\alpha_s}{8 \pi} G_{\mu \nu} \tilde{G}^{\mu \nu} + \frac{E}{N} \frac{a}{f_a} \frac{\alpha_{\rm em}}{8 \pi} F_{\mu \nu} \tilde{F}^{\mu \nu}  +\frac{\partial_\mu a}{2 f_a} \overline{f}_i \gamma^\mu \left( C^V_{ij} + C^A_{ij} \gamma_5 \right) f_j \, ,
\end{align}
with
\begin{align}
C^V_{u_i u_j} & =   \frac{C^{u_R}_{ij} + C^{u_L}_{ij}}{2 N} \, , & C^A_{u_i u_j} & =  \frac{ C^{u_R}_{ij} - C^{u_L}_{ij}}{2 N}  \, ,
\label{CVCA}
\end{align}
and analogous for the down-quark coupling. 
 
 \subsection{Scalar Potential}
 Compared to the SM the Yukawa Lagrangian has an extra $U(1)_h^n \times U(1)_\phi$ global symmetry that needs to be broken to a single $U(1)_{\rm PQ}$ factor by adding $n$ couplings in the scalar sector. Since  $U(1)_{\rm PQ} \ne U(1)_\phi$, we need at least one coupling of $\phi$. In general we can couple $h_i^\dagger h_j$ to an operator ${\cal O}_{ij} \in \{ \phi, \phi^*, \phi^2, \phi^{*2}\}$ at the renormalizable level. This determines the charge difference of the Higgs fields in terms of a free parameter $A_{ij}$  that can take the values $A_{ij}  \in \{ \pm 1, \pm 2\} $. Without loss of generality, we can therefore add suitable couplings in the scalar potential, which determine all Higgs charges in term of a charge $X_0$ of $h_0$ and $n-1$ parameters $A_i$, which can take discrete values that increase with $i$
\begin{align}
X_{i} & = X_0 + A_i X_\phi \, , & A_i & = \pm 1, \pm 2, \hdots, 2 i \, , & i & = 0, 1, \hdots , n-1 \, , 
\label{X12fix}
\end{align}
and $A_0 = 0$. Finally the charge $X_0$ is determined by requiring that the axion is orthogonal to the Goldstone eaten up by the $Z$-boson, which gives the conditions
\begin{align}
0 & =  \sum_{i} X_{i} \frac{v_{i}^2}{v^2}  \, , & 1 & =  \sum_{i}  \frac{v_{i}^2}{v^2}  \, ,
\end{align}
where the sum is taken over all Higgs fields with vev $v_i$ and $v = 246 \GeV$ denotes the electroweak vev. This gives
\begin{align}
X_0 & =  - X_\phi \sum_{i} A_i   \frac{v_{i}^2}{v^2}  \, , 
\end{align}
 which makes all Higgs charges and thus fermion charges to depend on continuous parameters, that is, the vacuum angles. As the anomaly coefficients are topological in nature, they have to be integers nevertheless, and indeed it is obvious that $2N$ in Eq.~\eqref{N} only depends on Higgs charge differences, so $X_0$ drops out, and up to an overall charge normalization $X_\phi$ the color anomaly coefficient is solely determined by $A_i$. Similarly for the electromagnetic anomaly  the charged lepton contribution $E_L$ will provide a complete cancellation of the $X_0$ dependence, and in the final ratio $E/N$
 the charge normalization $X_\phi$ cancels out leaving a rational number. Also in  fermion couplings $X_\phi$ cancels out, but these couplings depend  in general  on $X_0$. However, the combination 
 $C^{A}_{u_i u_j} + C^{A}_{d_i d_j}$ only depends on Higgs charge differences, so again does not depend on vacuum angles.  
 
 \subsection{Nucleophobic Models}
 We now restrict for simplicity to at most 3 Higgs doublets. This implies that the consistency conditions
 \begin{align}
 X_{h^u_{33}}   - X_{h^u_{a3}}  = X_{h^u_{3a}} - X_{h^u_{ab}} = X_{h^d_{a3}} - X_{h^d_{33}}   = X_{h^d_{ab}} - X_{h^d_{3a}}  \, .
\end{align}
 can only be fulfilled if in each equation Higgses are pairwise identified, giving 2 possibilities for each equation. Counting all distinct possibilities, one finds 7 distinct models, which are 
\begin{gather}
\label{sevenmodels}
\begin{pmatrix} h_b  h_b \\ h_a  h_a \end{pmatrix}  \begin{pmatrix} h_a  h_a \\ h_b  h_b \end{pmatrix} \, ,
\begin{pmatrix} h_a  h_a \\ h_a  h_a \end{pmatrix}  \begin{pmatrix} h_b  h_c \\ h_b  h_c \end{pmatrix} \, , 
\begin{pmatrix} h_b  h_a \\ h_b  h_a \end{pmatrix}  \begin{pmatrix} h_a  h_c \\ h_a  h_c \end{pmatrix}  \, ,
\begin{pmatrix} h_b  h_a \\ h_b  h_a \end{pmatrix}  \begin{pmatrix} h_c  h_a \\ h_c  h_a \end{pmatrix}  \, , \nonumber \\
\begin{pmatrix} h_a  h_b \\ h_a  h_b \end{pmatrix}  \begin{pmatrix} h_c  h_a \\ h_c  h_a \end{pmatrix}  \, ,
\begin{pmatrix} h_a  h_b \\ h_a  h_b \end{pmatrix}  \begin{pmatrix} h_a  h_c \\ h_a  h_c \end{pmatrix}  \, ,
\begin{pmatrix} h_b  h_c \\ h_b  h_c \end{pmatrix}  \begin{pmatrix} h_a  h_a \\ h_a  h_a \end{pmatrix}   \, .
\end{gather}  
 We also restrict to models that potentially have suppressed couplings to nucleons, which mainly depend on the valence quark couplings $C_u$ and $C_d$, given by 
 \begin{align}
C_u & \equiv C^A_u  = \frac{1}{2N} \left[ X_{h^u_{ab}} +   \left( X_{h^u_{33}} - X_{h^u_{3a}} \right) \xi^{u_{R}}_{11}  +   \left(  X_{h^u_{3a}} - X_{h^u_{ab}}  \right) \xi^{u_{L}}_{11} \right] \, , \\
C_d & \equiv C^A_d = \frac{1}{2N} \left[  -  X_{h^d_{ab}} +   \left(    X_{h^d_{3a}} - X_{h^d_{33}} \right)\xi^{d_{R}}_{11} + \left(  X_{h^d_{ab}} - X_{h^d_{3a}}  \right)  \xi^{d_{L}}_{11} \right] \, .
 \end{align}
For simplicity we restrict in the following to models without quark flavor violating, meaning either $\xi_{11} = 0$ or $\xi_{11} = 1$, which we take for the $u-$ and $d$-sector uniformly.  Couplings to nucleons are suppressed for $1 = C_u + C_d $, which implies for $\xi_{11} = 0$ models
\begin{align}
X_{h^u_{ab}} +  X_{h^u_{33}} -  X_{h^d_{ab}}  - X_{h^d_{33}}    =    0 \, , \qquad (\xi^{u_L}_{11} = \xi^{u_R}_{11}  = \xi^{d_L}_{11}  = \xi^{d_R}_{11}  = 0)
\end{align} 
and for  $\xi_{11} = 1$ models
\begin{align}
 X_{h^u_{ab}} -   X_{h^d_{ab}}      =     0  \, , \qquad (\xi^{u_L}_{11} = \xi^{u_R}_{11}  = \xi^{d_L}_{11}  = \xi^{d_R}_{11}  = 1) \, .
\end{align} 
The condition for $\xi_{11} = 0$ can again only be satisfied if Higgses are pairwise identified. Since we also want $2N = 2 X_{h^u_{ab}} +  X_{h^u_{33}} - 2 X_{h^d_{ab}}  - X_{h^d_{33}}$   to be non-zero for the QCD axion, we finally arrive at the following necessary conditions for nucelophobia in models with $n\le3$:
\begin{align}
X_{h^u_{ab}} & = X_{h^d_{33}} \ne X_{h^u_{33}} =  X_{h^d_{ab}}    \, , \qquad (\xi^{u_L}_{11} = \xi^{u_R}_{11}  = \xi^{d_L}_{11}  = \xi^{d_R}_{11}  = 0) \\
 X_{h^u_{ab}} & =   X_{h^d_{ab}}   \wedge  X_{h^u_{33}} \ne  X_{h^d_{33}}    \, , \qquad (\xi^{u_L}_{11} = \xi^{u_R}_{11}  = \xi^{d_L}_{11}  = \xi^{d_R}_{11}  = 1) \, ,
\end{align} 
so that the contribution to the color anomaly effectively comes only from a single family, $2N = X_{h^u_{ab}}  - X_{h^d_{ab}} $ for $\xi_{11} = 0$ and $2N = X_{h^u_{33}}  - X_{h^d_{33}} $ for $\xi_{11} = 1$ (cf. Ref.~\cite{DiLuzio:2017ogq}). 

Comparing to the possible 3HDMs in Eq.~\eqref{sevenmodels}, we see that for $\xi_{11} = 0$ only two structures allow for nucleophobia
\begin{gather}
{\rm Q2:} \qquad y_u \sim \begin{pmatrix} h_a  h_a \\ h_b  h_b \end{pmatrix} \, , \qquad y_d \sim \begin{pmatrix} h_b  h_b \\ h_a  h_a \end{pmatrix} \, , \nonumber \\
{\rm Q3:} \qquad y_u \sim \begin{pmatrix} h_a  h_b \\ h_a  h_b \end{pmatrix}  \, , \qquad y_d \sim \begin{pmatrix} h_b  h_a \\ h_b  h_a \end{pmatrix}  \, ,
\end{gather}  
with $a \ne b$, while for $\xi_{11} = 1$ there are also only  two structures
\begin{gather}
{\rm Q1:} \qquad y_u \sim  \begin{pmatrix} h_a  h_a \\ h_a  h_a \end{pmatrix}   \, , \qquad y_d \sim \begin{pmatrix} h_a  h_b \\ h_a  h_b \end{pmatrix} \, , \nonumber \\
{\rm Q4:} \qquad y_u \sim  \begin{pmatrix} h_c  h_a \\ h_c  h_a \end{pmatrix}   \, , \qquad y_d \sim \begin{pmatrix} h_c  h_{b}  \\ h_c  h_{b} \end{pmatrix}  \label{Q4}\, ,
\end{gather}  
where $a \ne b$ and $c \ne a$ (otherwise Q4 = Q1). The notation is chosen such that in all models 
\begin{align}
2N =&  X_a - X_b \, , & C_u & = \frac{X_a}{X_a - X_b} \, , & C_d & = \frac{- X_b}{X_a - X_b} \, . 
\end{align}
This makes manifest that indeed $C_u + C_d = 1 $, while the other condition for nucleophobia, $C_u - C_d = \frac{1}{3}$, requires for all models
\begin{align}
\frac{1}{3} = \frac{X_a + X_b}{X_a - X_b} \, .
\end{align}
For 2HDMs,  without loss of generality $h_a = h_1$ and $h_b = h_c = h_0$ , so that nucleophobia fixes the value of $X_0$ as
\begin{align}
X_0 & = - X_\phi A_1 \frac{v_1^2}{v^2} \overset{!}{=}  - \frac{1}{3} A_1 X_\phi   \, , 
\end{align}
with $v_1 \equiv s_\beta v, v_0 \equiv c_\beta v, $ and thus nucleophobia is achieved for $s_\beta^2 \approx 1/3$. Note that the numerical value of $A_1$ is unphysical as it is equivalent to re-defining the PQ charge normalization. Choosing e.g. $A_1 = 1$ (corresponding to the operator $h_1^\dagger h_2 \phi$) and $X_\phi = 1$, one obtains $X_1 = c_\beta^2$ and $X_0 = - s_\beta^2$, and we recover the four nucleophobic models proposed in Ref.~\cite{DiLuzio:2017ogq}, although we follow the notation in Ref.~\cite{Badziak:2021apn}. The resulting predictions are summarized in Table \ref{Qmodels2}.  

\begin{table}[t]
\centering
\begin{tabular}{|c||c||c|c||c|c|}
\hline
Model & $E_Q/N$  & $C^A_{u_i u_i}$ & $C^A_{d_i d_i}$ & $C^{V,A}_{u_i \ne u_j} $ & $C^{V,A}_{d_i \ne d_j} $   \\
\hline 
Q1 &  $2/3 + 6 c_\beta^2$ & $c_\beta^2 $ & $\xi^{d_R}_{ii} - c_\beta^2$ & 0 & $\xi^{d_R}_{ij}$ \\ \hline
Q2 &$-4/3 + 6 c_\beta^2$ & $c_\beta^2 - \xi^{u_L}_{ii}  $ & $-\xi^{d_L}_{ii} + s_\beta^2$ & $\pm \xi^{u_L}_{ij}$ & $ \pm\xi^{d_L}_{ij}$  \\ \hline
Q3 &  $- 4/3 + 6 c_\beta^2$ & $c_\beta^2 - \xi^{u_R}_{ii}$ & $-\xi^{d_R}_{ii} + s_\beta^2 $ & $- \xi^{u_R}_{ij}$ & $ - \xi^{d_R}_{ij}$  \\ \hline
Q4 & $-10/3 + 6 c_\beta^2$ & $- s_\beta^2 + \xi^{u_R}_{ii} $ & $s_\beta^2$ & $ \xi^{u_R}_{ij}$ & 0  \\
\hline
\end{tabular}
\caption{Axion couplings in the four nucleophobic 2HDMs Q1-Q4, as a function of the flavor parameters $\xi^{q_P}_{ij}$ and the vacuum angle $c_\beta \equiv \cos \beta, s_\beta \equiv \sin \beta$. Here $E_Q$ denotes the contribution of the quark sector to the electromagnetic anomaly coefficient $E$, to be added to the contribution $E_L$ from the charged lepton sector. In all models the domain wall number is trivial, $2N = 1$. Nucleophobia is achieved for $C_u \approx 2/3, C_d \approx 1/3$. \label{Qmodels2} }
\end{table}

In 3HDMs  one has the possibility to enforce nucleophobia by making $X_0 \ll 1$ (instead of 
$X_0 \approx -1/3$), so that fermion couplings can be made small by coupling them to $h_0$. This is because one can choose $h_a = h_1$, $h_b = h_2$ with $A_1 = 2, A_2 = -1$ (corresponding to the operators $h_1^\dagger h_0 \phi^2, h_2^\dagger h_0 \phi^\dagger  $)
\begin{align}
 \frac{X_a + X_b}{X_a - X_b}  = \frac{2 X_0/X_\phi + A_1  + A_2 }{A_1 - A_2} = \frac{1}{3} + \frac{2}{3} \frac{X_0}{X_\phi} \, .
\end{align}
With
\begin{align}
\frac{X_0}{X_\phi} = - \sum_{i} A_i   \frac{v_{i}^2}{v^2}  = \frac{v_{2}^2}{v^2} -  2  \frac{v_{1}^2}{v^2} \, .
\label{X0}
\end{align}
and parametrizing the vevs as $v_0 = s_2 v ,  v_1 =  c_1 c_2 v , v_2 = s_1 c_2 v$, one can choose vacuum angles such that $X_0/X_\phi = (1- 3 c_1^2) c_2^2 \ll1$. How small $X_0$ can be only depends on constraints from perturbativity of Yukawa couplings. In models Q1-Q4 the strongest bounds come from SM top and bottom Yukawas, which are given by $y_t = y_t^{\rm 3HDM} c_1 c_2, y_b = y_b^{\rm 3HDM} s_1 c_2$, where $ y_{b,t}^{\rm 3HDM}$ denote the couplings in the 3HDM, which neglecting running effects are bounded by  $ y_{b,t}^{\rm 3HDM} < \sqrt{16 \pi/3} \approx 4.1$. For more details and the resulting constraints on nucleophobia see Ref.~\cite{Bjorkeroth:2019jtx}. The constraints from perturbativity are relaxed in model Q5, which is obtained from the Q4 structure in Eq.~\eqref{Q4} by choosing $h_c = h_0$ (while the 3HDM model Q4 is obtained from setting $h_c = h_b = h_2$). This is because only $y_u$ and $y_d$ are proportional to $c_2$, while all other Yukawas are controlled by $s_2$, so that $c_2$ can be chosen much smaller than in Q1-Q4 without being in colnflict with perturbative unitarity. The resulting predictions for all models are summarized in Table \ref{Qmodels3}, upon choosing $X_\phi = 1$.

\begin{table}[t]
\centering
\begin{tabular}{|c||c||c|c||c|c|}
\hline
Model & $E_Q/N $  & $C^A_{u_i u_i}$ & $C^A_{d_i d_i}$ & $C^{V,A}_{u_i \ne u_j} $ & $C^{V,A}_{d_i \ne d_j} $   \\
\hline
Q1 &  $14/3 + 2 X_0$ & $2/3 + X_0/3$ & $- 2/3 - X_0/3 + \xi^{d_R}_{ii}$ & 0 & $\xi^{d_R}_{ij}$ \\ \hline
Q2 &$8/3 + 2 X_0$ & $2/3 + X_0/3 - \xi^{u_L}_{ii}$ & $1/3 - X_0/3 - \xi^{d_L}_{ii}$ & $\pm \xi^{u_L}_{ij}$ & $ \pm\xi^{d_L}_{ij}$  \\ \hline
Q3 &  $8/3 + 2 X_0$ & $2/3 + X_0/3 - \xi^{u_R}_{ii}$ & $1/3 - X_0/3 + \xi^{d_R}_{ii}$ & $- \xi^{u_R}_{ij}$ & $ - \xi^{d_R}_{ij}$  \\ \hline
Q4 & $2/3 + 2 X_0$ & $-1/3 + X_0/3 + \xi^{u_R}_{ii}$  & $1/3 - X_0/3$ & $ \xi^{u_R}_{ij}$ & 0  \\ \hline
Q5 & $2 + 2 X_0$ & $X_0/3 + 2/3 \xi^{u_R}_{ii}$ & $- X_0/3 + 1/3 \xi^{d_R}_{ii}$ & $ 2/3 \xi^{u_R}_{ij}$ & $ 1/3 \xi^{d_R}_{ij}$   \\
\hline
\end{tabular}
\caption{Axion couplings in the five potentially nucleophobic 3HDM models Q1-Q5, as a function of the parameters $\xi^{q_P}_{ij}$ and $X_0$. Here $E_Q$ denotes the contribution of the quark sector to the electromagnetic anomaly coefficient $E$, to be added to the contribution from the charged lepton sector. In all models the domain wall number is $2N = 3$ and $X_0 \ll 1$. Nucleophobia is achieved for $C_u \approx 2/3, C_d \approx 1/3$. \label{Qmodels3} }
\end{table}
\subsection{Lepton Yukawa Sector}
The general charged lepton Yukawa Lagrangian is given by  
\begin{align}
{\cal L} & =  y^{e}_{33} \overline{\ell}_{L3} e_{R3} \tilde{h}^e_{33} + y^{e}_{3a}  \overline{\ell}_{L3} e_{Ra} \tilde{h}^e_{3a} + y^{e}_{a3} \overline{\ell}_{La} e_{R3} \tilde{h}^e_{a3} + y^{e}_{ab} \overline{\ell}_{La} e_{Rb} \tilde{h}^e_{ab} + {\rm h.c.}  \, , 
\label{Llep}
\end{align}
in the same notation as Eq.~\eqref{L1}. We begin by restricting to a {\bf 2}+{\bf 1} flavor structure, denoting general Higgs couplings as
\begin{align}
y_e \sim \begin{pmatrix} h^e_{ab}  h^e_{a3} \\ h^e_{3a}  h^e_{33} \end{pmatrix}  \, ,
\end{align}
which fixes charged lepton charges in terms of Higgs charges as
\begin{align}
X_{\ell_a} & = X_{h^e_{33}} - X_{h^e_{a3}} + X_{\ell_3} \, , &
X_{e_a} & = X_{h^e_{3a}}  + X_{\ell_3} \, , & X_{e_3} & = X_{h^e_{33}} + X_{\ell_3} \, ,
\end{align}
and gives a single consistency condition
\begin{align}
X_{h^e_{a3}} - X_{h^e_{33}}   = X_{h^e_{ab}} - X_{h^e_{3a}} \, .
\end{align}
The contribution to the electromagnetic anomaly coefficient $E_L$ is given by 
\begin{align}
E_L & = - 2 X_{h^e_{ab}} - X_{h^e_{33}} \, ,  
\end{align}
and axion couplings to charged leptons in the mass basis read
\begin{align}
C^{e_L}_{ij} & = \left( X_{h^e_{a3}} - X_{h^e_{33}}   - X_{\ell_3}  \right) \delta_{ij} -   \left( X_{h^e_{a3}} - X_{h^e_{33}}   \right) \xi^{e_{L}}_{ij}    \, , \nonumber \\
C^{e_R}_{ij} & = \left(   - X_{h^e_{3a}} - X_{\ell_3}  \right)  \delta_{ij} + \left(  X_{h^e_{3a}} - X_{h^e_{33}}  \right)  \xi^{e_{R}}_{ij}   \, .
\end{align}
Restricting to at most 3HDMs, again the consistency condition can only be pairwise satisfied, giving 2 possible structures, which are universal LH or RH charged lepton charges
\begin{gather}
y_e \sim \begin{pmatrix} h_d  h_e \\ h_d  h_e \end{pmatrix} \quad [ER] \qquad \qquad {\rm or } \qquad\qquad y_e \sim \begin{pmatrix} h_d  h_d \\ h_e  h_e \end{pmatrix}  \quad [EL] \end{gather}  
We are interested in suppressed electron couplings, which read
\begin{align}
C_e \equiv C^A_e = \frac{1}{2N} \left(  - X_d \delta_{ij} +  \left( X_d - X_e   \right) \begin{cases} \xi^{e_{R}}_{11} & {\rm ER} \\ \xi^{e_{L}}_{11} & {\rm EL} \end{cases}  \right)
\end{align}
In 2HDM, where both Higgs charges are large, the only way to suppress $C_e$ is by tuning $\xi_{11}$. There are 2 possibilities for identifying $h_{d}$ and $h_e$ with $h_0$ and $h_1$, giving 
\begin{align}
{\rm E1:}  \qquad y_e & \sim  \begin{pmatrix} h_1  h_1 \\ h_0  h_0 \end{pmatrix}   \, , \nonumber \\
 {\rm E2:}  \qquad y_e & \sim  \begin{pmatrix} h_0  h_0 \\ h_1  h_1 \end{pmatrix}    \, .
\end{align}  
where we restricted to EL for simplicity, which gives identical predictions for axion couplings as ER, upon replacing $\xi^{e_L}_{ij} \leftrightarrow \xi^{e_R}_{ij} $. One finally obtains for 2HDMs the predictions in Table~\ref{Lmodels2}. Note that these models cannot be simultaneously electro- and muon- phobic if the conditions for nucleophobia and suppressed LFV couplings in the $\mu$-$e$ sector are imposed.

\begin{table}[t]
\centering
\begin{tabular}{|c||c||c||c|c|c||c|c|c||}
\hline
Model & $E_L/N$  & $C^A_{e_i e_i}$ & $C^{V,A}_{e_i \ne e_j} $   \\
\hline
E1 &  $2 - 6 c_\beta^2$ & $- c_\beta^2 + \xi^{e_L}_{ii}  $ & $\mp \xi^{e_L}_{ij}$ \\
E2 & $4 - 6 c_\beta^2$  & $ s_\beta^2 - \xi^{e_L}_{ii} $ & $ \pm\xi^{e_L}_{ij}$  \\
\hline
\end{tabular}
\caption{Axion couplings in the two potentially electrophobic 2HDM models EL1 and EL2, as a function of the parameters $\xi^{e_L}_{ij}$ and $c_\beta \equiv \cos \beta, s_\beta \equiv \sin \beta$. Predictions for ER1 and ER2 are identical, upon $\xi^{e_L}_{ij} \to \xi^{e_R}_{ij}$ and in the last column $\mp \xi^{e_L}_{ij} \to  \xi^{e_R}_{ij}, \pm \xi^{e_L}_{ij} \to  - \xi^{e_R}_{ij}$. Here $E_L$ denotes the contribution of the charged lepton sector to the electromagnetic anomaly coefficient $E$, to be added to the contribution from the quark sector. Electrophobia is achieved for $\xi^{e_L}_{11} \approx c_\beta^2$ (E1) or $\xi^{e_L}_{11} \approx  s_\beta^2$ (E2) . \label{Lmodels2} }
\end{table}

In 3HDMs, one can suppress the electron coupling simply by choosing $h_d = h_0$ ($\xi_{11} = 0$) or $h_e = h_0$ ($\xi_{11} = 1$), since $X_0$ has small PQ charge. Then there are  in total 5 distinct choices (upon L $\leftrightarrow$ R)
\begin{align}
{\rm E1:} \qquad y_e \sim  \begin{pmatrix} h_0  h_0 \\ h_1  h_1 \end{pmatrix}   \, ,  \nonumber \\
{\rm E2:} \qquad y_e \sim  \begin{pmatrix} h_0  h_0 \\ h_2  h_2 \end{pmatrix}   \, ,  \nonumber \\
{\rm E3:} \qquad y_e \sim  \begin{pmatrix} h_0  h_0 \\ h_0  h_0 \end{pmatrix}   \, ,  \nonumber \\
{\rm E4:} \qquad y_e \sim  \begin{pmatrix} h_1  h_1 \\ h_0  h_0 \end{pmatrix}   \, ,  \nonumber \\
{\rm E5:} \qquad y_e \sim  \begin{pmatrix} h_2  h_2 \\ h_0  h_0 \end{pmatrix}   \, .  
\end{align}  
They gives rise to the models in Table~\ref{Lmodels3}. Note that E1 and E2 can be simultaneously electro- and muonphobic in the absence of lepton flavor violation ($\xi_{33} = 1$), while in E3 there is no LFV and all lepton couplings are small.
\begin{table}[t]
\centering
\begin{tabular}{|c||c||c||c|c|c||c|c|c||}
\hline
Model & $E_L/N$  & $C^A_{e_i e_i}$ & $C^{V,A}_{e_i \ne e_j} $   \\
\hline
E1 &  $-4/3 -2 X_0 $ & $-X_0/3 -2/3 \xi^{e_L}_{ii}  $ & $\pm 2/3 \xi^{e_L}_{ij}$ \\ \hline
E2 & $2/3 -2 X_0 $  & $-X_0/3 + 1/3 \xi^{e_L}_{ii}  $ & $ \mp 1/3 \xi^{e_L}_{ij}$  \\ \hline
E3 & $-2 X_0 $  & $ -X_0/3   $ &  0 \\ \hline
E4 & $-8/3 -2 X_0 $  & $-2/3 -X_0/3 + 2/3 \xi^{e_L}_{ii}   $ & $ \mp 2/3 \xi^{e_L}_{ij}$  \\ \hline
E5 & $4/3 -2 X_0 $  & $ 1/3 -X_0/3 -1/3 \xi^{e_L}_{ii}   $ & $ \pm 1/3 \xi^{e_L}_{ij}$  \\ 
\hline \hline
E6 & $-2/3 -2 X_0 $  & $  - X_0/3  - \delta_{i3} /3    $ & $  \pm \sqrt{2}/3 \left( \delta_{i2}  \delta_{j3} +  \delta_{i3}  \delta_{j2} \right) $  \\
\hline
\end{tabular}
\caption{Axion couplings in the six potentially electrophobic 3HDM models EL1-EL6, as a function of the parameters $\xi^{e_L}_{ij}$ and $X_0$. Predictions for ER1-ER6 are identical, upon $\xi^{e_L}_{ij} \to \xi^{e_R}_{ij}$ and in the last column $\mp \xi^{e_L}_{ij} \to  \xi^{e_R}_{ij}, \pm \xi^{e_L}_{ij} \to  - \xi^{e_R}_{ij}$ and $\pm \sqrt{2}/3 \to - \sqrt{2}/3$ for ER6. Here $E_L$ denotes the contribution of the charged lepton sector to the electromagnetic anomaly coefficient $E$, to be added to the contribution from the quark sector. One can check that $E_L/N = 2 \sum_i C^A_{e_i e_i}$ Electrophobia is achieved for $X_0 \ll 1$ in E3 and additionally $\xi^{e_L}_{11} \approx 0$ (E1,E2) or $\xi^{e_L}_{11} \approx 1$ (E4,E5). The model E6 is special as all three leptons carry different PQ charges. \label{Lmodels3} }
\end{table}
Interestingly, for the 3HDM there are three combinations that are also photo-phobic, since they have $E/N = 2$, namely Q1E4, Q4E5, and Q5E3. In particular in model Q5E3 also all lepton couplings are small, so that there are sizable couplings only to first generation quarks, and couplings  to nucleons and pions are also suppressed. This avoids essentially all phenomenological constraints, apart from mild SN1987A constraints from nucleon couplings. As discussed above, this also implies that constraints on vacuum angles from perturbative unitarity are very mild, thus allowing to achieve $X_0 \ll 1$ through $c_2 \ll 1$. 

Finally we relax the assumption of a {\bf 2}+{\bf 1} flavor structure and consider the possibility that in the charged lepton sector each lepton carries different PQ charge. We however still restrict to flavor universal charges for either LH or RH leptons, needed to satisfy the consistency constraints. Thus we consider two scenarios: 

\begin{gather}
y_e \sim \begin{pmatrix} h_d  h_e h_f \\ h_d  h_e h_f \\ h_d  h_e h_f \end{pmatrix} \quad [ER] \qquad \qquad {\rm or } \qquad\qquad y_e \sim \begin{pmatrix} h_d  h_d h_d \\ h_e  h_e h_e \\ h_f h_f h_f \end{pmatrix}  \quad [EL] \end{gather}  
This gives only models different than those discussed previously, if all three Higgses are different. Therefore the contribution $E_L$ is fixed, and given by
\begin{align}
E_L = - X_d - X_e - X_f = - (3X_0  + A_1 + A_2) = -3 X_0  - 1 \, , 
\end{align}
and 
\begin{align}
E_L/N =  - \frac{2}{3} \left( 3 X_0  + 1 \right) = -2 X_0 - 2/3 \, . 
\end{align}
Combining this model with Q2 or Q3 therefore gives a model with suppressed photon couplings. In addition we can simultaneously suppress electron and muon couplings. The charged lepton couplings read 
\begin{align}
(C^A_e)_{ij} = - \frac{1}{2N}  (V^*_{EP})_{ki}  (V_{EP})_{kj} X_{k} \, , 
\end{align}
where $P = L/R$ for EL/ER and $X_k = \{ X_{d}, X_e, X_f \}$. We now want to get suppression of the $C_e, C_\mu$ and $C_{e \mu}$ in order to avoid stringent constraints from WDs, SN1987A and LFV searches. This can indeed be achieved for choosing $h_d = h_0, h_e = h_1, h_f = h_2$, so
\begin{align}
{\rm E6:} \qquad y_e \sim \begin{pmatrix} h_0  h_0 h_0 \\ h_1  h_1 h_1 \\ h_2  h_2 h_2 \end{pmatrix}
\end{align}
for the LH model, the RH model is analogous. This gives
\begin{align}
(C^A_e)_{ij} = - \frac{1}{3}  \left[ X_0 \delta_{ij} - (V^*_{EL})_{3i}  (V_{EL})_{3j} + 2 (V^*_{EL})_{2i}  (V_{EL})_{2j}  \right] \, . 
\end{align}
Now choosing $V_{EL}$ to be just a rotation in the 2-3 sector, 
\begin{align}
V_{EL} = \begin{pmatrix} 1 & & \\ & c_e & s_e \\ &  -s_e & c_e \end{pmatrix}\, , 
\end{align}
with $s_e =  \sqrt{2/3}, c_e = \sqrt{1/3}$, one finds
\begin{align}
(C^A_e)_{ij} = - \frac{1}{3}  X_0 \delta_{ij} - \frac{1}{3} \begin{pmatrix} 0 & 0 & 0 \\ 0 & 0 & \sqrt{2} \\ 0 & \sqrt{2} & 1 \end{pmatrix} \, , 
\end{align}
so that diagonal and off-diagonal couplings in the E6 model read 
\begin{align}
(C^A_e)_{ii} & = - \frac{1}{3}  X_0  - \frac{1}{3} \delta_{i3}  \, , & (C^{V,A}_e)_{i\ne j} & = \pm \frac{\sqrt{2}}{3} \left( \delta_{i2}  \delta_{j3} +  \delta_{i3}  \delta_{j2} \right) \, ,
\end{align}
and for $X_0 \ll1$ only $C_\tau$ and $C_{\tau \mu}$ are non-vanishing.

\clearpage
\bibliographystyle{JHEP} 
\bibliography{Bibliography}

\end{document}